\documentclass{aastex631}

\usepackage{comment}

\newcommand{\vhel}{v_{hel}}
\newcommand{\vunb}{v_{\infty}}
\newcommand{\vexcess}{v_{+}}
\def\las{\mathrel{\hbox{\rlap{\hbox{\lower3pt\hbox{$\sim$}}}\hbox{\raise2pt\hbox{$<$}}}}}
\def\gas{\mathrel{\hbox{\rlap{\hbox{\lower3pt\hbox{$\sim$}}}\hbox{\raise2pt\hbox{$>$}}}}}

\submitjournal{the Astrophysical Journal, accepted 21 March 2025}

\begin{document}

\title{An upper limit on the interstellar meteoroid flux at video sizes from the Global Meteor Network}

\author[0000-0002-1914-5352]{Paul Wiegert}
\author[0009-0005-7887-1505]{Vanessa Tran}
\author[0000-0001-8927-7708]{Cole Gregg}
\author[0000-0003-4166-8704]{Denis Vida}
\author[0000-0001-6130-7039]{Peter Brown}
\affiliation{Department of Physics and Astronomy \\
The University of Western Ontario \\
London, Canada}
\affiliation{Institute for Earth and Space Exploration (IESX) \\
The University of Western Ontario \\London, Canada}

%% Note that the \and command from previous versions of AASTeX is now
%% depreciated in this version as it is no longer necessary. AASTeX 
%% automatically takes care of all commas and "and"s between authors names.

%% AASTeX 6.31 has the new \collaboration and \nocollaboration commands to
%% provide the collaboration status of a group of authors. These commands 
%% can be used either before or after the list of corresponding authors. The
%% argument for \collaboration is the collaboration identifier. Authors are
%% encouraged to surround collaboration identifiers with ()s. The 
%% \nocollaboration command takes no argument and exists to indicate that
%% the nearby authors are not part of surrounding collaborations.

%% Mark off the abstract in the ``abstract'' environment. 
\begin{abstract}

Material arriving at our solar system from the Galaxy may be detected at Earth in the form of meteors ablating in our atmosphere. Here we report on a search for interstellar meteors within the highest-quality events in the Global Meteor Network (GMN) database. No events were detected that were conclusively hyperbolic with respect to the Sun; however, our search was not exhaustive and examined only the top 57\% of events, with a deeper examination planned for future work.

This study's effective meteoroid mass limit is $6.6\pm0.8 \times 10^{-5}$~kg (5~mm diameter at a density of 1000~kg~m$^{-3}$). Theoretical rates of interstellar meteors at these sizes range from 3 to 200 events globally per year. The highest rates can already be largely excluded by this study, while at the lowest rates GMN would have to observe for 25 more years to be 50\% confident of seeing at least one event. GMN is thus well positioned to provide substantial constraints on the interstellar population at these sizes over the coming years. This study's results are statistically compatible with a rate of interstellar meteors at the Earth at less than 1 per million meteoroid impacts at Earth at mm sizes, or a flux rate of less than $8 \pm 2 \times 10^{-11}$~km$^{-2}$~h$^{-1}$ at the 95\% confidence level.
\end{abstract}

%% Keywords should appear after the \end{abstract} command. 
%% The AAS Journals now uses Unified Astronomy Thesaurus concepts:
%% https://astrothesaurus.org
%% You will be asked to selected these concepts during the submission process
%% but this old "keyword" functionality is maintained in case authors want
%% to include these concepts in their preprints.

%% From the front matter, we move on to the body of the paper.
%% Sections are demarcated by \section and \subsection, respectively.
%% Observe the use of the LaTeX \label
%% command after the \subsection to give a symbolic KEY to the
%% subsection for cross-referencing in a \ref command.
%% You can use LaTeX's \ref and \label commands to keep track of
%% cross-references to sections, equations, tables, and figures.
%% That way, if you change the order of any elements, LaTeX will
%% automatically renumber them.
%%
%% We recommend that authors also use the natbib \citep
%% and \citet commands to identify citations.  The citations are
%% tied to the reference list via symbolic KEYs. The KEY corresponds
%% to the KEY in the \bibitem in the reference list below. 

\section{Introduction} \label{sec:intro}

An object with a heliocentric speed above 42.135~km~s$^{-1}$ at 1 au from the Sun is on an escape orbit, unbound from our star, and this is often taken as a proxy for its being of interstellar origin \citep{Hajdukova_2019}.  More precisely, the heliocentric speed must be more than the parabolic velocity $\vunb$ at a particular heliocentric distance $R$ given by
\begin{equation}
\vunb(R) = \sqrt{\frac{2GM_{\odot}}{R}} = 42.135 \left(\frac{R}{1~{\rm au}} \right)^{-1/2}~{\rm km~s}^{-1} \label{eq:escapespeed}
\end{equation}
where $G$ is the universal gravitational constant and $M_{\odot}$ is the Sun's mass.  In the case of a meteor observed in the Earth's atmosphere, if its speed prior to falling into Earth's gravitational well was greater than $\vunb$, it is possible that it originated outside our Solar system, a so-called 'interstellar' meteor. The value of $\vunb$ is 42.135~km~s$^{-1}$ at Earth's nominal distance from the Sun of 1 au. However, because Earth's orbit is slightly eccentric, its heliocentric distance ranges from 0.983 to 1.067 au during the course of the year and thus $\vunb$ varies from $41.79 \leq \vunb \leq 42.50$~km~s$^{-1}$.

Meteors with observed heliocentric speeds above $\vunb$ may also be produced from within our solar system. Beta meteoroids \citep{zoober75, coomuloli93} are initially bound particles that are accelerated onto unbound orbits by solar radiation pressure. However, these are only small (less than one micron in size) particles, not visible as meteors in Earth's atmosphere by either video or radar meteor detection instruments. A gravitational slingshot from a planet may also place a bound object of arbitrary size onto a hyperbolic orbit but such events are expected to be rare. At Earth, one in $10^4$ optical meteors is expected to fall into this category but these are typically traveling only 100 m~s$^{-1}$ above the heliocentric escape velocity at Earth \citep{wie14} and hence difficult to distinguish from bound particles for typical measurement precision. Thus particles above micron-size within our solar system found to be on unbound orbits are likely to be interstellar, in the absence of a recent encounter with a planet. 

A particle arriving at the solar system with an excess speed $\vexcess$ will accelerate as it falls into the Sun's gravitational potential and will have a heliocentric speed $v(R) = \sqrt{\vunb(R)^2+ \vexcess^2}$.  The velocity dispersion $\sigma$ of stars near the Sun within the disk of the Milky Way is about 20~km~s$^{-1}$ \citep{del65, mihbin81, bintre87} and so interstellar meteoroids are expected to have typical excess speeds $\vexcess \approx \sigma$, and to reach the Earth 
$v(R=1~\rm{au}) \approx 47 \rm{~km/s~at~1~au}$. 
As a result, interstellar meteors are expected to arrive at Earth with speeds a few~km~s$^{-1}$ above $\vunb$, making them relatively easy to distinguish from the abundant nearly-unbound meteors produced within our solar system, at least if sufficiently accurate and precise speed measurements can be made.

Here we use the Global Meteor Network (GMN) \citep{vidseggur21} database of meteor events to search for meteoroids moving above the hyperbolic limit with respect to the Sun. Because of measurement uncertainties, meteors with speeds near the hyperbolic limit are often incorrectly assigned an interstellar origin when they are actually on bound solar system orbits. This usually occurs when measurement uncertainties are underestimated, or when the uncertainties straddle the border between bound and unbound trajectories with the nominal speed being unbound while the true value is bound \citep{fis28,hajpau07,hajstewie20}.

This paper reports on our first examination of the GMN database of optical meteors, where we extract only the highest quality events in a search for true interstellar meteors. A secondary goal is to develop an understanding of the random and systematic measurement errors inherent in the GMN data set. Any real-world system must cope with a complex and changing environment containing clouds, aurora and other environmental effects as well as aircraft, satellites and other spurious man-made sources. Other both common and rare potentially confounding effects include  meteors near the detection limit, meteors near the horizon or the edge of the camera field of view, multiple simultaneous or fragmenting meteors, miscalibration, improper automated meteor picks, poor triangulation geometry.  Future efforts will push deeper into the GMN database as our toolkit and understanding of the system capabilities increase, but this study is focused on GMN's most reliably measured events.

\subsection{Previous studies}
%%%%%
Detection of interstellar meteoroids has been a controversial topic in meteor science for more than a hundred years \citep{Stohl_1970}. Prior to the advent of instrumental measurements, it was widely accepted on the basis of visual observations that many unbound (presumably interstellar in origin) meteoroids encountered the Earth \citep{Opik_1950}. For example, a catalog of visually determined orbits by \cite{vonhof25} contains 79\% hyperbolic orbits \citep{fis28}. However, the advent of regular photographic multi-station velocity estimates and even more precise radar (head echo) velocities in the mid-20th century showed that few, if any, reliably measured heliocentric meteoroid velocities exceeded the parabolic limit \citep{Almond_1950}. It was largely accepted for the next half century that essentially all meteoroids with apparently excess velocity were the result of measurement errors \citep{Hajdukova_1994, hajstewie19}. In the case of photographic and radar observations, the measurement errors in question arise from a combination of underestimation of the statistical uncertainty together with systematic limitations arising from pixel size, frame rate and other system limitations.  Some events  will by chance have poor multi-station geometry and this is a major factor in measurement uncertainty for photographic and radar velocity vectors. Uncertainty can also arise from the analysis process itself, for example when the atmospheric deceleration of a meteor prior to its being detected must be computed or estimated.

The first unambiguous detection of interstellar particles (ISPs) in
our solar system was made by the Ulysses and Galileo spacecraft
\citep{Baguhl1995}. However, the detected particles were on the order
of 1 $\mu$m in radius (10$^{-11}$ g), below the meteor detection limit
for almost all instruments at the Earth. Interstellar meteoroids were
reported to have been detected through transverse scattering
measurements of meteor ionization trails by the Advanced Meteor Orbit
Radar (AMOR) \citep{Baggaley_2000} in the mid-1990s. However,
subsequent analysis of over 20 million radar meteors detected by the
Canadian Meteor Orbit Radar (CMOR) \citep{Froncisz_2020} have failed
to confirm this initial detection, albeit at larger sizes. Much of the
uncertainty they encounter comes from the difficult process of
correcting for atmospheric drag on these small particles in order to
extrapolate meteor velocities back to their pre-atmospheric values,
but the main limitation for radar multi-static estimates of velocity
is station – specular scattering point geometry, in particular
geometries where the specular points from multiple stations all lie at
nearly the same location on the meteor trail. Detection of
interstellar meteoroids from either radial or transverse scattering
radars remains controversial \citep[e.g.][]{Hajdukova_2019}.

Optical measurements, which often can be more precise than radar and therefore are well suited to searching for real interstellar particles (ISPs), have also been used to survey for interstellar meteoroids. Dedicated early surveys using optical cameras showed few, if any, apparently hyperbolic meteors \citep{McKinley_1961} resolving as measurement error the apparent paradox of large numbers of hyperbolic meteors first reported from visual observations \citep{lov54}.

More recent optical studies have much larger numbers than the comparatively small early surveys. For example, \citet{Hajdukova__2014} examined a catalog comprised of 64,650 video meteors observed in Japan between 2007 and 2008. Of these detections, 7,489 appeared to have hyperbolic orbits. After filtering for meteors with low errors in measured velocity and rejecting meteors associated with showers, 238 retained hyperbolic orbits - these were all attributed to being the result of measurement error. They found that "the vast majority of hyperbolic meteors in the database have definitely been caused by inaccuracy in the velocity determination".

In addition to the search for hyperbolic meteors captured in the course of broader surveys, \citet{Musci_2012} performed a dedicated optical survey for mm-sized interstellar meteoroids with emphasis on accurately estimating uncertainties. Out of 1739 double station meteor measurements, they found only two meteors as being unbound beyond 3$\sigma$ of the estimates uncertainties. They concluded that "Detailed examination leads us to conclude that our few identified events are most likely the result of measurement error" but do not give the specific cause.   

The discovery of a large hyperbolic object, 1I/‘Oumuamua, in 2017 \citep{Meech_2017} has reinvigorated interest in the quest for interstellar meteoroids. This was the first confirmation of a macroscopic interstellar object in the solar system. Presently, together with comet 2I/Borisov, these remain the only large interstellar interlopers whose orbits are well enough determined to be unambiguous examples of ISPs. 

That smaller, mm-sized ISPs are transiting the solar system remains virtually certain \citep{Murray2004}. It is difficult to imagine a selection filter which would remove ISPs at such sizes. However, the observational evidence for this population remains scant. Reported detections to date remain unconvincing when considering their associated measurement uncertainties particularly of the velocity. For example, \cite{Hajdukova_2019} find that "the vast majority of hyperbolic meteors in the [\cite{son09} optical meteor] database have definitely been caused by inaccuracy in the velocity determination". It remains to be shown that a meteoroid in the mm-cm sized range is clearly unbound from our solar system given proper account of its associated uncertainty. The task requires a huge time-area atmosphere collecting product to allow sufficient statistics to overcome the apparently low ISP flux in these ranges, which is  estimated at $6 \times 10^{-9}$ to $4 \times 10^{-7}$ meteoroids km$^{-2}$~yr$^{-1}$ at masses $\gas 7 \times 10^{-5}$~kg (5 mm diameter at a density of 1000 kg~m$^{-3}$, Section~\ref{sec:masslimit}), or 3 to 200 events globally per year (Section~\ref{sec:expectedflux}).

%%%%%

\section{Methods} \label{sec:methods}

\subsection{Global Meteor Network}
The Global Meteor Network (GMN) \citep{vidseggur21} is an international collaborative initiative dedicated to the continuous monitoring and analysis of meteors and meteorite-dropping fireballs, long-term characterization of meteor showers, determination of meteoroid flux and size distributions, and enhancement of public awareness regarding near-Earth meteoroid environments. It is structured as a network of over 1000 low-cost, highly sensitive CMOS video cameras deployed across more than 40 countries. Each camera station employs open-source meteor detection software running on Raspberry Pi computers, enabling efficient data collection and processing. The metric data are uploaded to a central server which automatically computes meteor trajectories and their associated uncertainties using the Monte Carlo trajectory solver developed by \citet{vida2020theory, vida2020results}. 

Since its inception in December 2018, the network has amassed millions of precise meteoroid orbits which are publicly available\footnote{GMN orbital data set: \url{https://globalmeteornetwork.org/data/} (accessed November 1, 2024)} under the CC-BY-4.0 license. Our sample was obtained from the GMN public database for dates beginning on 2018 December 10 and ending 2024 October 5. 

\subsection{Filtering}

\begin{longtable}{p{13cm} p{2cm}}

\caption{Selection Parameters for Trajectory Calculation and Processing  \label{tab:initialfilters} }\\ 

\textbf{Parameter} & \textbf{Value/Range} \\ 
\endfirsthead
\hline
\textbf{Parameter} & \textbf{Value/Range} \\ \hline
\endhead

\hline
\endfoot

\multicolumn{2}{c}{\textbf{Trajectory Calculation Parameters}} \\ \hline
Minimum number of measurement points per observation & 4 \\ 
Maximum time difference between meteor observations from different stations & 10.0 seconds \\ 
Minimum and maximum distance between stations & 5 - 600 km \\ 
Minimum convergence angle \(Q_c\) & 3.0 degrees \\ 
Maximum number of stations included in the trajectory solution & 8 \\ 
Max difference between velocities from different stations & 25\% \\ 
Minimum and maximum average in-atmosphere velocities & 3 - 73~km~s$^{-1}$ \\ 
Minimum and maximum begin height & 50 - 150 km \\ 
Minimum and maximum end height & 20 - 130 km \\ 
Maximum angle between radiants for merging candidate trajectories & 15 degrees \\ 

\multicolumn{2}{c}{\textbf{Trajectory Filtering Parameters}} \\ \hline
Stations with radiant residuals larger than this fixed amount are removed & 180 arcseconds \\ 
Stations with radiant residuals below the value will be retained, regardless of other filters & 30 arcseconds \\ 
If 3+ stations, any with angular residuals more than this many times the mean are removed & 2.0 \\ 

\multicolumn{2}{c}{\textbf{Post-Processing Parameters}} \\ \hline
Minimum number of points on the trajectory for the station with the most points & 6 \\ 
Minimum convergence angle & 10 degrees \\ 
Maximum radiant error & 1.0 degrees \\ 
Maximum geocentric velocity error & 1\% \\ 
Maximum begin height & 160 km \\ 
Minimum end height & 60 km \\

\end{longtable}

A full trajectory solution requires a meteor be detected by at least two stations. The automated heuristic deciding which observations from individual stations to pair into a candidate meteor event applies strict filters to reject false positives and incorrect pairings, and chooses the best set of observations to form a trajectory by rejecting sub-optimal stations. The quality filtering mainly relies on meteor observations forming tight trajectories (the average fit residuals from individual stations not exceeding 3 arcminutes) and the observed velocities between sites matching within several percent \citep{vidseggur21}. 
The filters applied by GMN are empirically
determined from extensive experience manually reducing individual GMN events. Since each camera's observing conditions, orientation and geometrical relationship to other cameras nearby is unique, we recognize that no single set of cut threshold values necessarily works for all situations, but the general character of the cuts is to remove events with insufficient or poor quality data. As an example, the first cut in Table~\ref{tab:initialfilters} requires four measurements per observation. As the meteor peaks above the noise threshold, the first and the last frame will always only have a partial track (the rest is below the noise threshold), so requiring four frames means that the middle two frames are guaranteed to be 100\% complete.  All the selection parameters are hard-coded into the GMN processing pipeline and cannot be varied as part of this study. They are described in more detail in \cite{vidzubseg16,  vida2020results, vida2020theory, vidseggur21} and are listed in Table~\ref{tab:initialfilters} for reference. Events that did not meet these criteria do not proceed to further filtering steps. A total of 1.59 million individual GMN meteor events (73\%) passed this first step out of an initial 2.18 million orbits computed by GMN. 

This automated pipeline provides a first list of potential ISPs based on the observed meteor's parameters, but it can potentially yield erroneous results in a small number of cases. It may not correctly interpret fragmenting meteors, meteors with persistent trains, multiple meteors occurring simultaneously within the same field of view, or it may produce a lower-quality solution due to various environmental or weather phenomena. As a result, we will use the automated results for the initial selection of potential interstellar meteors, but will ultimately rely on the manual reduction of the event by a human operator to provide the most accurate determination of a prospective ISP-meteor trajectory.

\subsubsection{Initial Filtering} \label{initialfiltering}

To isolate interstellar candidate meteors, the data was filtered to remove all events with heliocentric speeds $\vhel < 42$~km~s$^{-1}$ and whose radiant coincided with a known active meteor shower. The events were also required to be at least 5 sigma above the 42~km~s$^{-1}$ threshold, that is $(\vhel - 42) / \sigma > 5$, where $\sigma$ is the uncertainty in the heliocentric speed provided by the trajectory solver. To increase the overall quality of events in the next stages, we also tightened some thresholds above and beyond those set by the GMN database and listed in Table~\ref{tab:initialfilters}. Specifically the convergence angle ($Q_c$) was required to be at least 20 deg, and the maximum begin heights and minimum end heights were required to be 160 km and 60 km respectively. These last conditions impose some assumptions on the physical properties of interstellar meteors which may not be correct if their composition is quite different from those of solar system meteoroids. This assumption is necessary to reduce the number of false positives but events outside this range will be examined more closely in future work.    

The next two final filters are less critical and chosen to extract the highest-quality events.  First, though the heliocentric speed of interstellars is necessarily high, their in-atmosphere speed need not be. The Earth's orbital velocity of 30~km~s$^{-1}$ is added vectorially to the particle's heliocentric velocity and could increase or decrease its in-atmosphere speed, spreading a 42~km~s$^{-1}$ heliocentric speed across a range of in-atmosphere speeds from 12~km~s$^{-1}$ to 72~km~s$^{-1}$. As events with high in-atmosphere speeds are less precisely measured owing to the limited number of observed frames, the initial in-atmosphere speed of events that would pass to the next stage was required to be less than 50~km~s$^{-1}$. Additional justification for this step will be provided in Section~\ref{uncertainty}. This reduces our sample to 2122 events.

\begin{figure}
\plotone{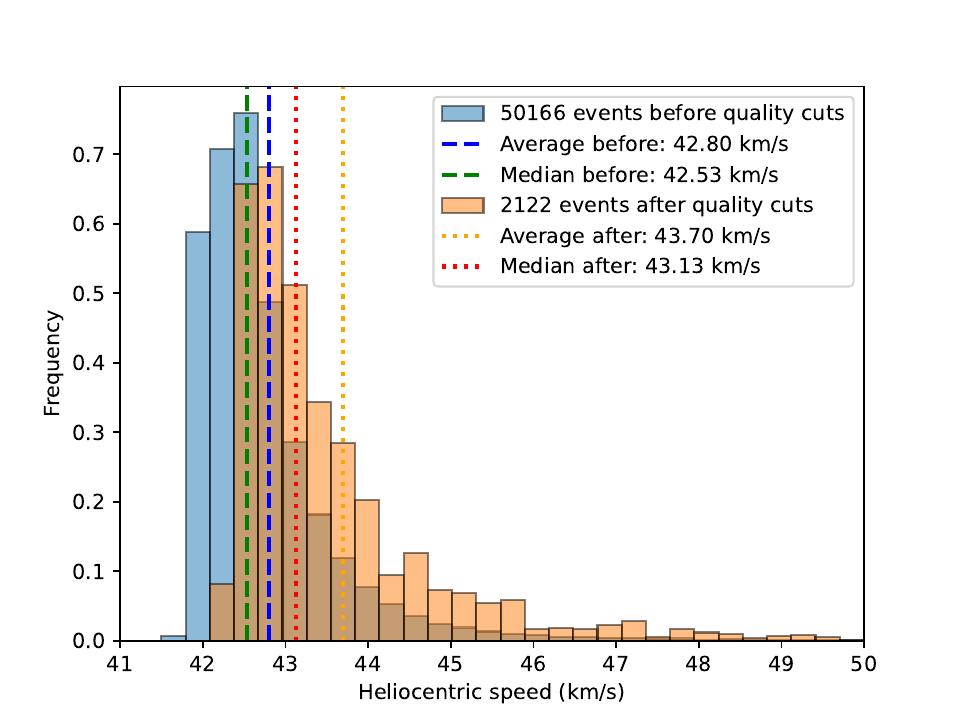}
\caption{The velocity distributions of hyperbolic GMN events before and after the quality cuts we impose. \label{fig:vh_before_and_after}}
\end{figure}

These conditions are velocity-dependent and do result in a change in the overall velocity distribution. This can be seen in Figure~\ref{fig:vh_before_and_after} where the heliocentric speed distribution of unbound events is shown both before and after the cuts of the previous paragraph. Many events near the hyperbolic limit are removed, and the peak shifts to slightly higher speeds. The number of events to examine is reduced by about a factor of 25 from 50166 to 2122 without excessive distortion of the distribution itself.

Second, we required that the heliocentric speed be at least 44~km~s$^{-1}$ to avoid cases too close to the 42~km~s$^{-1}$ boundary, a step which significantly reduced the number of interstellar candidates to a manageable level. In total, these two filters reduce our overall sample size to 472 high-quality interstellar candidates.

\subsubsection{Stage 2 - Pre-Reduction Manual Review} \label{prereduction}

For the 472 interstellar candidates, the
automated GMN solution is manually inspected to determine whether the solution is reliable. The applied quality criteria are as follows:
\begin{itemize}
\item Metric Measurement Residuals: all residuals should be consistently less than 100 arcseconds.
\item Systematic Measurement Residuals: their mean deviation should be within one standard deviation of their scatter.
\item Perspective Angle: for a two-station solution, the meteor’s trajectory should not be going directly towards/away from a camera; the ideal scenario is if the meteor is located between two cameras with a trajectory perpendicular to the camera FOV: the perspective angles must $> 20^o$ and $< 75^o$ or  $> 110^o$. 
\item Lags:  The deceleration profile represented by the lags (namely, how far the observed position differs from that predicted by a constant speed solution computed from the first few frames of the event) should not differ by more than 10\% between stations.
\item The individual station velocities should be within 10\% of each other and the automated solution.
\end{itemize}

If an automated solution passes most or all of the requirements above, the raw data files are downloaded, in preparation for a manual reduction process: 34 events reach this stage. The raw data files must then pass through a set of checks detailed below.

\subsubsection{Stage 3 - Pre-Reduction Camera Manual Review}

Before manual reduction, events are checked that 
\begin{itemize}
% \item The meteor can be seen in all cameras  
\item At least one camera captured the entire meteor, and others have at least 5-6 frames where it is clearly visible 
\item The meteor has stars near it to enable a good astrometric solution 
\item The times reported by all the cameras line up with each other – there is a possibility for the event to actually be two nearly simultaneous but separate events conflated with each other
\item There are no dropped frames in the meteor video
\item The meteor is not passing through any significant obstructions, clouds/weather/environmental effects
\end{itemize}

Events which pass these criteria were then manually reduced to give the most accurate orbital determination. Only 13 events reach the stage of being manually reduced in detail.

\subsection{Manually reduced events}
A full manual reduction of a meteor event observed by GMN cameras begins with careful re-calibration of astrometric plates on stars visible on the same video as the meteor. The positions of the stars on the image are measured by fitting a 2D Gaussian point spread function. They are then manually paired with catalog stars, upon which a 7th order radial polynomial distortion model is fit, also including a correction for lens anisotropy \citep{vidseggur21}. The fit is manually refined by taking care to remove any outliers and making sure there are no systematic trends in the fit. At least 30-40 stars are selected, making sure that the whole field of view is equally covered and that there are stars in the area around the meteor. For the most common GMN cameras with 4~mm lenses, an average astrometric fit residuals between 0.5-1.0~arcmin are usually achieved. Photometric calibration includes extinction and vignetting correction, achieving fit residuals between 0.1-0.2~mag.

Metric measurements of meteor positions on individual video frames are performed by manual centroiding on the meteor head for each meteor frame. The trajectory solution is computed continuously as picks at each station are made, allowing checking of the quality of the trajectory as individual data points are added to the solution. The process finishes once all video frames which show the meteor are measured. In some cases the measurements are further refined based on the trajectory fit graphs (both fit residuals and the dynamics) and outliers are rejected.

After manual reduction, the trajectory parameters and the orbit are recomputed and their uncertainties are estimated by computing 100 Monte Carlo realizations, providing the final orbital values and uncertainties we present herein following the procedure described in \cite{vida2020theory}.

\section{Discussion and Results}

All manually reduced events were found to be consistent with bound heliocentric orbits, though many had error bars which extended nominally into the unbound regime. To examine their significance, we computed the probability that $v_{hel}$ was above the hyperbolic limit at the Earth at the time in question assuming the errors are Gaussian distributed. None were found to have a probability larger than 49\% of being on a unbound orbit,  and the probability of any event having a $v_{hel}$ above the 47~km~s$^{-1}$ expected for true interstellars is less than 4\%. See Table~\ref{tab:reducedevents} for details.

\begin{deluxetable*}{|c|c|c|c|c|c|c|c|}
\tablewidth{0pt}
\tablecaption{The GMN events that were manually reduced. The columns under $\vhel$ indicate the heliocentric speed of the event from the automatically and manually generated solutions. The final manually-obtained heliocentric eccentricity is in column $e$. $R_{\oplus}$ is the Earth's heliocentric distance at the time of the event and $v_{\infty}$ the corresponding heliocentric escape speed (Eqn~\ref{eq:escapespeed}). The final columns gives the probability that the event is unbound, and of having $\vhel > 47$~km/s expected for true interstellars, under the simple assumption that the errors are Gaussian. Probabilities less than $10^{-9}$ are listed as zero. \label{tab:reducedevents}}
\tablehead{
\colhead{Identifier} &  \multicolumn{2}{c}{$v_{hel}$ (km/s)} & 
\colhead{$e$} & \colhead{$R_{\oplus}$} & 
\colhead{$v_{\infty}$} & 
\multicolumn{2}{c}{Prob. of}  \\
\colhead{} & \colhead{auto} & \colhead{manual} & 
\colhead{} & \colhead{(au)} & \colhead{(km s$^{-1}$)} & \colhead{$\vhel>v_{\infty}$} & \colhead{$\vhel > 47$~km/s}
}
\startdata
20191105\_074753.938371 &  50.42 & 39.0 $\pm$ 2.2 & 0.91 $\pm$ 0.06 & 0.992 & 42.302 & 0.063 & $1 \times 10^{-4}$\\
20200305\_092152.467353 &  44.15 & 41.8 $\pm$ 1.2 & 0.95 $\pm$ 0.11 & 0.992 & 42.301 & 0.325 &  $5 \times 10^{-6}$\\
20200411\_065019.471331 & 45.60 & 41.7 $\pm$ 3.0 & 0.97 $\pm$ 0.16 & 1.002 & 42.085 & 0.448 & $4 \times 10^{-2}$\\
20200714\_104743.729656 &  46.79 & 37.89 $\pm$ 0.03 & 0.911 $\pm$ 0.001 & 1.017 & 41.779 & 0.000 & 0 \\
20200925\_072128.279662 &  49.01 & 39.1 $\pm$ 1.6 & 0.92 $\pm$ 0.03 & 1.003 & 42.066 & 0.029 & $2 \times 10^{-7}$\\
20220225\_113605.069662 &  49.03 & 41.76 $\pm$ 0.23 & 0.978 $\pm$ 0.008 & 0.990 & 42.344 & 0.005 & 0\\
20220411\_204649.331547 &  49.66 & 41.98 $\pm$ 0.14 & 0.995 $\pm$ 0.008 & 1.002 & 42.085 & 0.236 & 0\\
20220430\_044938.390913 &  45.98 & 41.27 $\pm$ 0.52 & 0.937 $\pm$ 0.045 & 1.007 & 41.979 & 0.085  & 0\\
20221214\_191339.014896 & 44.11 & 42.24 $\pm$ 0.21 & 0.983 $\pm$ 0.017 & 0.984 & 42.457 & 0.153 & 0\\
20230124\_004120.554640 &  44.02 & 41.00 $\pm$ 0.53 & 0.925 $\pm$ 0.027 & 0.984 & 42.459 & 0.003 & 0\\
20230628\_062714.788236 &  46.86 & 37.83 $\pm$ 0.088 & 0.773 $\pm$ 0.004 & 1.017 & 41.778 & 0.000  & 0\\
20231215\_214203.653783 &  44.29 & 42.45 $\pm$ 0.40 & 0.9999 $\pm$ 0.006 & 0.984 & 42.458 & 0.493 & 0\\
20240113\_204657.410119 &  44.13 & 42.146 $\pm$ 0.074 & 0.972 $\pm$ 0.007 & 0.983 & 42.475 & 0.000 & 0\\
\enddata
\end{deluxetable*}

Though great care was taken to filter the GMN data for the highest-quality meteors, the occurrence of rare and unforeseen events is inevitable in such a large sample.  Below we discuss some of the subtle reasons we found that some GMN meteors appeared interstellar on first examination but proved otherwise upon careful review. This provides a sense of the sort of unusual real-world issues which other meteor networks should expect to encounter when searching for ISPs.

\begin{itemize}
\item  For a few months after they were initially set up, a few GMN cameras in Korea were running with incorrect frame rates of 12.5 frames per second instead of 25. This resulted in a number of extremely clean events that appeared exactly as well-measured interstellar meteors. It was only by vigilant review of all the information about the cameras that these events, which proved to be bound solar system meteors once the correct frame rate was applied,  were identified. This mishap inadvertently provided a proof of concept that the GMN pipeline can process and flag interstellar meteors if they were to actually appear.

\item Another rare cause of apparently interstellar events is the coincidental detection of two meteors on apparently similar trajectories within a few seconds by the same set of cameras. Because of their spatial and temporal coincidence, the algorithm can pair them incorrectly,  which can result in a malformed solution with an incorrect velocity determination. GMN uses a 10 second correlation window, so similar events which occur within this time frame can be tentatively considered as a candidate trajectory and in rare cases pass all quality filters. Though 10 seconds is longer than the duration of most meteors, this window cannot easily be shortened in practice, as it allows for trajectories of longer meteors which were caught at different portions of their in-atmosphere trajectories to be computed. In addition, this also allows the inclusion of measurements from stations whose clocks may have drifted over time, in rare cases when when the Network Time Protocol (NTP) service malfunctions or the camera loses Internet connection.
 
\item A more mundane cause of interstellar misidentification results from events which are distant from one or more of the cameras or which have less than optimal observing geometry. The speed determination depends on a reliable localization of the meteor location within the video frame. The center-of-light of the meteor does not in all cases correspond with its center-of-mass due to the effects of fragmentation and wake. For brighter meteors and fireballs in particular, difficult observing geometry can result in sub-pixel shifts that can produce an artificial overall acceleration of the meteor during the brightening phase. Often in these cases the meteor may be consistent with both bound and unbound trajectories within the uncertainties, depending on which analyst performs the manual reduction. A tell tale sign of this issue is an apparent acceleration (much larger than that due to gravity) of the meteor while undergoing ablation.

\item A well-determined trajectory depends on the precise location of the meteor in the sky. This may be difficult to measure through thin or patchy cloud even if the meteor itself is clearly visible, as on-sky position is measured by reference to nearby stars which are used to compute an astrometric plate solution relating pixel position to celestial coordinates. Bad weather or bright aurora might prevent accurate measurement of even a bright meteor by reducing the number of stars visible in the images to the point of degrading the astrometry.
\end{itemize}

\subsection{Uncertainties in speed} \label{uncertainty}

\begin{figure}
%\plotone{GMN_kresak.pdf}
\plotone{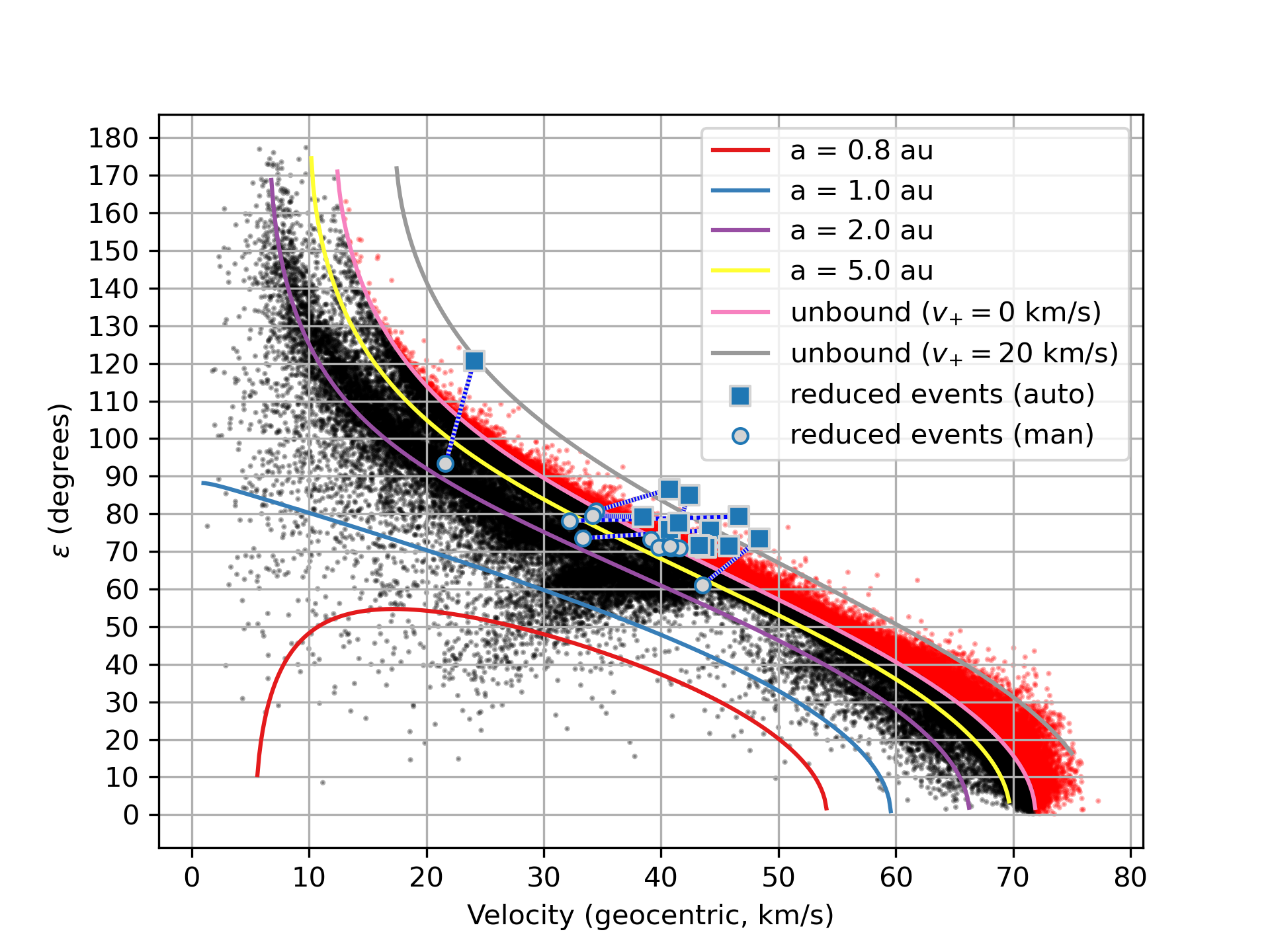}
\caption{The automatically determined geocentric speed versus the elongation $\varepsilon$ of the meteor's radiant from the Earth's apex. Lines of constant semimajor axis $a$ are superimposed, with a parabolic orbit indicated by an arrival speed $\vexcess = 0$~km/s, and an arrival speed expected for interstellar meteors by $\vexcess = 20$~km/s. Nominally hyperbolic events are indicated in red. Events with $a$ near 3~au have been decimated to improve plot visibility. The 13 candidates of Table~\ref{tab:reducedevents} are indicated with a dotted line connecting their automated speed with the manually reduced  speed. Note that these candidates cluster near the maximum geocentric speed (50~km~s$^{-1}$) examined in this study, while faster meteors were excluded. Higher geocentric speed meteors are typically seen in fewer video frames and will be examined in a follow-up study.\label{fig:kresak}}
\end{figure}

A plot of the geocentric speed of the meteoroid prior to falling into the Earth's gravitational well (traditionally known as the "geocentric speed" or $V_g$) versus the elongation $\varepsilon$ from the Earth's apex is shown in Figure~\ref{fig:kresak}. This type of figure was introduced by \cite{krekre76} and developed by \cite{hajstewie19,hajstobar24} to illustrate the effects of measurement uncertainties on meteor orbit determination. 
By superimposing upon the plot lines of constant semimajor axis $a$, the sensitivity of $a$ to measurement errors both in the speed and the radiant directions becomes evident. Where the lines of constant semimajor axis are closer together, (eg. along a diagonal in the middle of the plot and unfortunately also where most meteor events occur), is where measurement uncertainties either in speed or radiant direction can most easily result in a bound orbit being mistakenly identified as unbound. 

From the automated reduction pipeline, we find that 5.9\% of GMN events are initially identified as hyperbolic. This is consistent with what is typically seen in other surveys \citep{hajstewie20}, and is associated with the effect of measurement uncertainties. This can been seen qualitatively from Fig.~\ref{fig:kresak} where the majority of "hyperbolic" meteors (in red) are at high geocentric speeds, which have high in-atmosphere speeds and so tend to appear in only a few video frames. If these were true interstellars, we would expect them to appear all along the $\vexcess = 20$~km/s line in Fig.~\ref{fig:kresak} but this is not the case. The fact that the majority of apparently hyperbolic meteors appear towards the apex of the Earth's way (that is, at the lowest values of $\varepsilon$) at the highest atmospheric speeds and exactly where the measurement uncertainty is the largest alerts us to the danger of naively interpreting our data.

\begin{figure}
\plotone{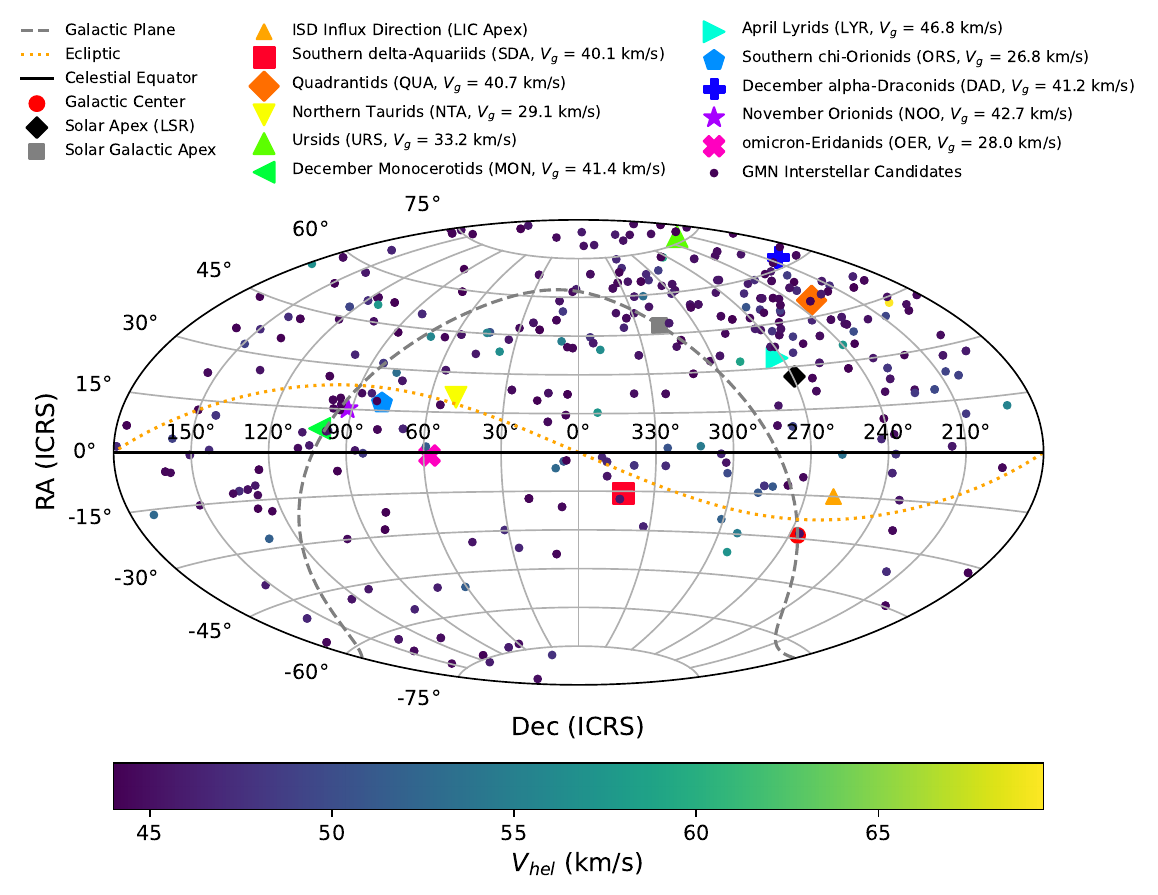}
\caption{The radiants in RA and Dec of the 472 fastest ($\vhel \geqq 44$~km~s$^{-1}$) events along with the radiants of some known meteor showers and some relevant fiducial directions such as the Galactic Center. For the showers, the 10 most active annual IAU meteors showers that are seen by GMN (that is, that contain significant mm sized particles) and that have in-atmosphere speeds below 50 km/s. Directional values: the position of the Galactic Center \citep{Reid_2004}; the Solar apex with respect to the Local Standard of Rest (LSR) \citep{jaschek_solar_1992}; the Solar apex with respect to the Galactic frame (Solar Galactic Apex); and the interstellar dust (ISD) influx direction into the Solar System as reported by spacecraft Cassini, Ulysses, and Galileo, due to the motion of the Solar System through the Local Interstellar Cloud (LIC) \citep{sterken_flow_2012}. No strong signature of the Galaxy or of meteor showers is seen.  \label{fig:onskyB}}
\end{figure}

A second way of looking at the data that is helpful is to examine the radiant directions of interstellar candidates and to see if there is any correlation with galactic reference directions. Figure \ref{fig:onskyB} shows such a plot of candidate ISPs in their relationship to the Galaxy and to known meteor showers. The plot shows the meteors with the highest automatically-generated $\vhel$ in Right Ascension vs Declination, along with known meteor showers and some galactic reference directions. There is no clear association with either meteor showers or the galaxy, which supports the conclusion that measurement errors of high-speed solar system meteors are the primary source of interstellar false positives.

\section{Computing the flux limit} \label{sec:masslimit}
None of the events examined in this work are convincingly interstellar. Some are consistent with both bound and unbound solutions, and many events have been deliberately rejected due to quality issues discussed earlier in Section~\ref{sec:methods}. Though this work does not provide an exhaustive examination of the GMN catalog, it is designed to isolate the most likely and best quality interstellar candidates. On this basis we conclude provisionally that the GMN sample is consistent with no interstellar meteors having been observed during its operations to date, for meteors with geocentric speeds $< 50$~km~s$^{-1}$. Future work is planned to examine the database in more detail as new events accumulate and as our tools for meteor analysis improve, but for the moment we will use the temporal and spatial parameters of the GMN survey to date to set limits on the flux of interstellar meteors into the Earth's atmosphere at visual sizes.

\subsection{Mass limit}
\begin{figure}
\plotone{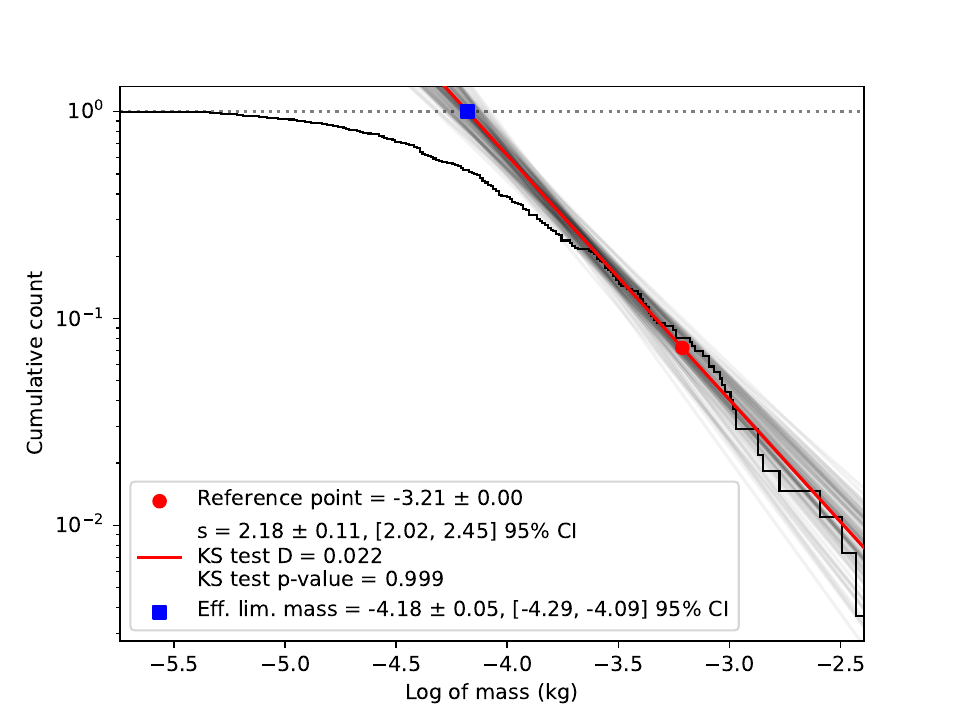}
\caption{The cumulative mass function for GMN detections examined in this work. Only the 273 events with geocentric speeds below 50 km/s and automatically determined heliocentric speeds above 42 km/s are included to provide the best estimate of our sensitivity to interstellar events. This work's effective mass limit is $10^{-4.18 \pm 0.05} = 6.6 \pm 0.8 \times 10^{-5}$~kg. See the text for more details.\label{fig:masslimit}}
\end{figure}

To estimate the limiting flux of interstellar meteors detected by GMN, we take the following approach. Since the cameras are all similar, for simplicity we adopt a single mean limiting magnitude for the entire system, with a corresponding mean limiting mass. The cumulative mass distribution for GMN is presented in Figure~\ref{fig:masslimit}. The masses used in the analysis were computed from GMN's photometric measurements and using a fixed luminous efficiency of $\tau = 0.7\%$. The cumulative mass index and the mass limit were computed using the method developed in \cite{vida2020draconids}. In short, the method first uses the Maximum Likelihood Estimation to fit a gamma distribution to the mass distribution. The inflection point where the slope starts decreasing on the upward leg of the gamma function is taken as the point at which events begin being lost due to sensitivity effects. The reference point, at which the mass distribution's slope is to be estimated, is taken to be 0.4 dex (corresponding to roughly one meteor magnitude) larger in mass than the inflection point. A line is fitted to the cumulative distribution at the reference point using the Kolmogorov-Smirnov test's p-value as the measure of goodness-of-fit. When extended upward to unity (see Fig.~\ref{fig:masslimit}) cross $y=1$ at the point typically taken to be the "effective limiting mass". This procedure essentially finds the mode of the probability density function of the size distribution \citep{blacamkin16,vida2020draconids}.

The GMN's average effective mass limit for candidate ISPs in this work with in-atmosphere velocities below 50 km/s is $6.6 \pm 0.8 \times 10^{-5}$~kg. This corresponds to a sphere of 5~mm in diameter at a density of 1000~kg~m$^{-3}$, and [10~mm, 2~mm] for densities of [100~kg~m$^{-3}$, 7800~kg~m$^{-3}$] corresponding to fluffy cometary aggregates through iron.

\subsection{Collecting area}
The multi-station collecting area of GMN has increased dramatically since its beginning in 2018 to over 1000 systems in 38 countries at the end of this analysis in October 2024. Since a full trajectory solution requires overlapping fields of view, the collecting area is a complicated function of time as new systems come online. However, since GMN collects information on the field of view (FOV) of each camera on a nightly basis, it is a simple task to examine the field of view of each camera in the network on each night, determine which portions overlap with nearby stations, and sum the resulting area to obtain the precise area-time product needed to determine a true flux.  Some care is needed to avoid multi-counting regions where 3 or more cameras overlap. 

\begin{figure}
\plotone{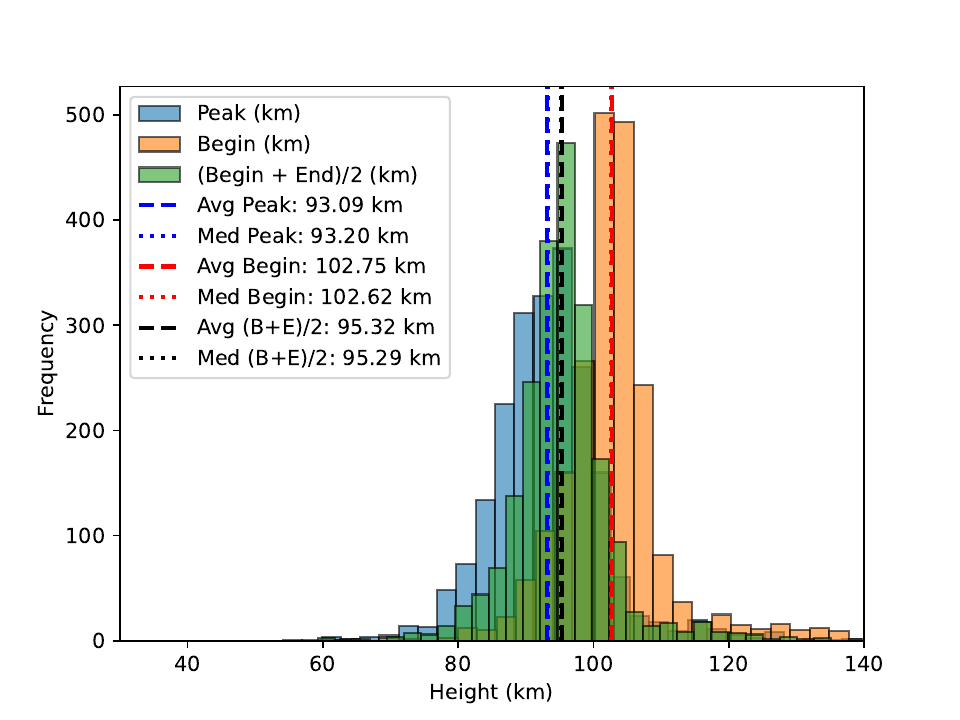}
\caption{Various measures of meteor height for the 2122 GMN events remaining after initial filtering (see Section~\ref{initialfiltering}.)  \label{fig:heights}}
\end{figure}

In practice, we proceed as follows. Each night, each stations produces a file that contains geographic coordinates (latitude, longitude) of vertices of a  polygon representing the camera’s FOV projected onto an altitude of 100 km. The 100 km altitude is where most GMN events are seen. Figure~\ref{fig:heights} shows the average and median begin heights (where meteors are first seen, 103 km), peak heights (where meteors are brightest, 93 km) and for (begin height + end height)/2 (traditionally used in flux computations, 95 km). Our flux estimates depend linearly on the collecting area, and we estimate our uncertainty in that quantity to be of order 10\%. 

In practice, each camera only overlaps with a few others, and doing a full binary search checking the overlap of each camera pair is prohibitive. To streamline the search for overlapping areas, each station is assigned to one of seven continents based on its country code. These are reprojected using \texttt{pyproj} \citep{alan_d_snow_2024_13864781} and \texttt{shapely.ops.transform} \citep{gillies_2024_13345370} from WGS84 to a suitable Universal Transverse Mercator (UTM) coordinate system near the centre of the continent and stored as a shapely polygon object. Each pair of FOV polygons on the same continent is then checked for overlap via the \texttt{shapely.intersection} function.  

\begin{figure}
\plotone{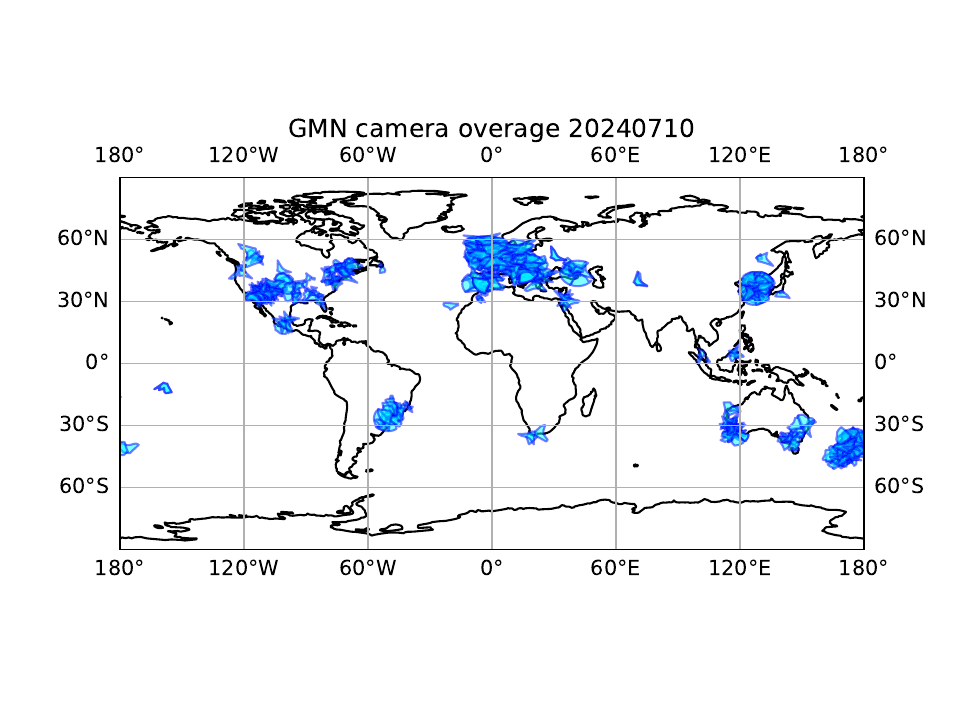}
\caption{The total single-camera area coverage of GMN on 2024 July 10.  \label{fig:coverage}}
\end{figure}

\begin{figure}
\plotone{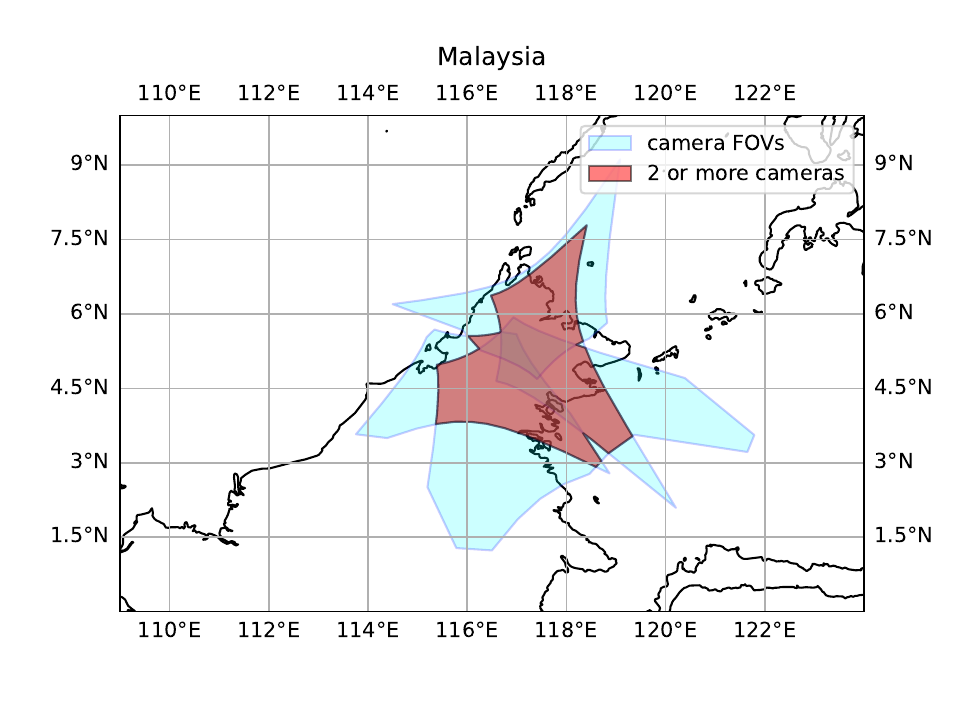}
\caption{The individual fields of view (in blue) of the 6 stations in Malaysia along with the area of overlap (in red) where full orbital solutions are possible. \label{fig:Malaysia}}
\end{figure}

From the list of overlapping FOVs, we produce a graph (in the mathematical "graph theory" sense) of the GMN network where each vertex is a camera station, and any camera stations with overlapping fields of view are connected vertices. On any given date the GMN network graph is then decomposed into its connected subgraphs, corresponding to a collection of stations which overlap with at least one other station, but are disconnected from the rest of the network. Figure~\ref{fig:coverage} shows single station FOVs for all GMN stations on July 10, 2024 to illustrate how the network separates into sub-networks. Stations in areas where GMN cameras are numerous like Europe are typically part of a large subgraph, but will be disconnected from distant subgraphs in (for example) Africa, Oceania or North America because there is no continuous GMN coverage over the oceans. In July 2024, GMN was composed of 21 individual subgraphs. Each subgraph has its collecting area computed using \texttt{shapely.ops.unary\_union}, which merges the multiple intersecting fields of view into a single polygon, removing any overlapping areas. An example subgraph showing the fields of view and overlapping regions of six cameras in Malaysia is shown in Fig~\ref{fig:Malaysia}.

\begin{figure}
\plotone{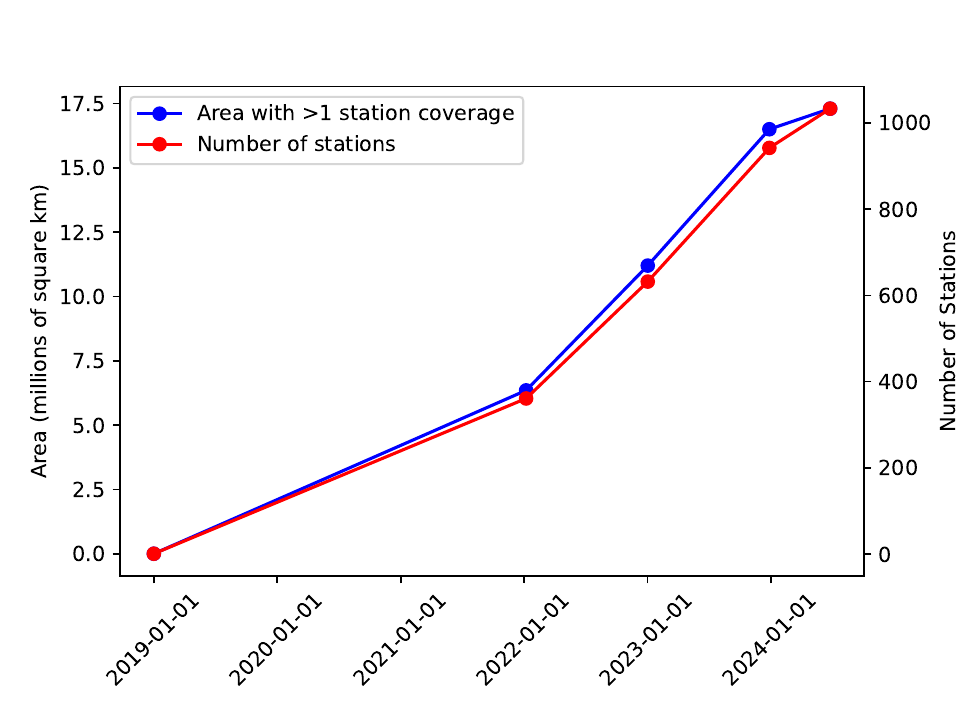}
\caption{The total overlapping area coverage of GMN along with the number of stations, as a function of time.  \label{fig:time-area}}
\end{figure}

The files describing the fields of view of the GMN cameras are only available after mid-2021, but we can extrapolate backwards to late 2018 with only minimal error, as most of the collecting area occurs at later dates due to the rapid growth of the network. Figure~\ref{fig:time-area} shows the total area of the sky at 100 km altitude observed by at least 2 cameras of the GMN as a function of time. As of mid-2024, the GMN covers a total of 17.5 million sq km by multiple cameras at the height of 100 km. This roughly translates to 3.5\% of the total Earth's area, or 12\% of the Earth's landmass.

\subsubsection{Flux limit}

\begin{figure}
\plotone{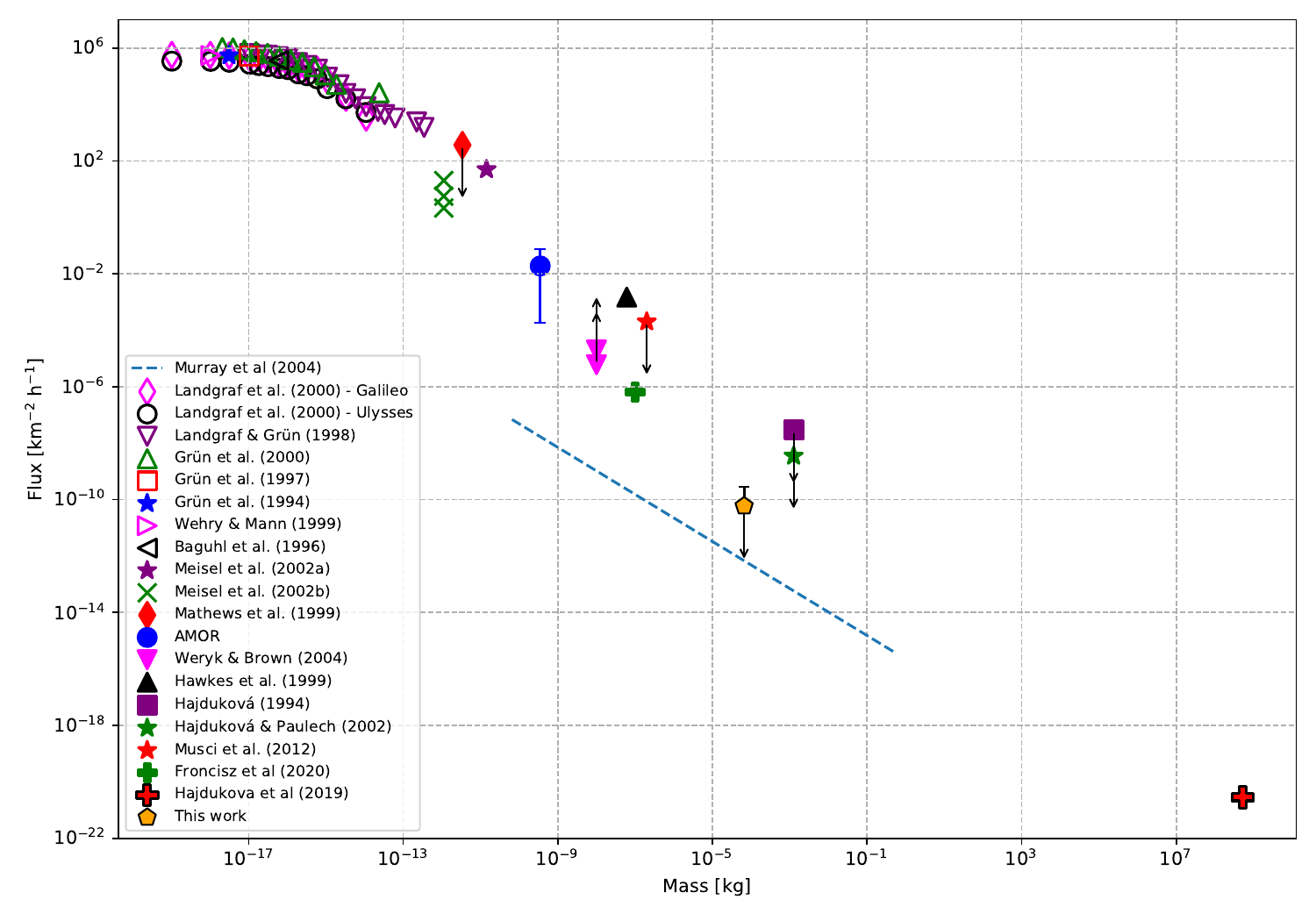}
\caption{The upper limit on the flux of interstellars at optical sizes from this study along with other values from the literature, adapted from \cite{Musci_2012}. The values listed as AMOR include those from \cite{bagtayste93, taybagste96, Baggaley_2000}. The value on the extreme right from \cite{hajstewie19} is based on 1I/'Oumuamua. The blue dashed line is a theoretical estimate of expected flux from young main-sequence star systems such as $\beta$ Pic from \cite{Murray2004}, from their nominal size of 50 microns in diameter up to 10 cm using their adopted differential size slope of -2.5. \nocite{langru98,lanbaggru00,grulanhor00,grustabag97,grugusman94,wehman99,baggrulan96,meijanmat02a,meijanmat02b,matmeijan99,werbro04,hawclowoo99,bag00,bagtayste93,taybagste96,haj94,hajpau02,Musci_2012,Froncisz_2020,hajstewie19, Murray2004} \label{fig:flux}}
\end{figure}

Given a non-detection of any interstellar events over the time period covered by this study, we compute a limit on the flux $F$ of interstellar meteoroids at these sizes of $F < \frac{1}{\phi A \cdot t}$ where $A \cdot t$ is our raw integrated time-area product for GMN, and $\phi$ is an efficiency factor. A value of $A \cdot t = 2.4 \pm 0.4 \times 10^{11}$~km$^2$~hr was computed by adding up the area of the atmosphere observed by GMN cameras, where we assume GMN cameras to be active on average 12 hours per day. Care was taken to track the growth of the network over time and compute the total overlapping areas for working cameras on a daily basis. To account for weather and other environmental effects, we simply assume an efficiency factor of $\phi = 0.45$ due to long-term average global cloud cover over the land of 55\% \citep{king2013spatial}.  To put it into context, since its inception and until mid-2024, the GMN's total effective time area product is roughly 10 global days. Other factors that affect our overall efficiency are our selection of only events with $v_g <50$~km/s (which include only 57\% of all events) as well as our inclusion of only higher quality events (which make up 73\%). Our net overall efficiency is thus $\phi = 0.45\times 0.57 \times 0.73 = 0.19$. For a Poisson process, the exact or 'conservative' 95\% confidence interval is given by 
\begin{equation}
[0.5 \chi^2(2N, \alpha/2),  0.5 \chi^2(2(N+1), 1-\alpha/2)]
\end{equation}
where $\chi^2$ is the Chi-Square critical value, $N$ is the number of events, and $\alpha$ is the significance level ($\alpha = 0.05$ for the 95\% confidence level) \citep{geh86, meehahesc17}. The numerical values when no events are detected ($N=0)$ are [0, 3.69], so our flux limit in this work $F < \frac{3.69}{\phi A \cdot t} = 8 \pm 2 \times 10^{-11}$~km$^{-2}$~h$^{-1}$ at the 95\% confidence level. Figure~\ref{fig:flux} shows this value relative to other flux values from the literature,. 

\subsection{Expected flux of interstellar particles at millimeter sizes} \label{sec:expectedflux}
There are only a few estimates of the expected fluxes of interstellar particles at video meteor sizes in the literature. \cite{Murray2004} estimate the fluxes at Earth of particles from Asymptotic Giant Branch (AGB) stars, Young Stellar Objects (YSOs) and young main-sequence stars. AGB stars are not expected to produce much material at mm-sizes, their dust being primarily at micron sizes and below \citep{hof08,mathof11,vanwaldan20}. YSOs may produce millimeter sized grains, which could be entrained in bipolar jets and outflows and escape their parent system, but \cite{Murray2004} conclude that none of the YSOs currently in the Sun's vicinity are at the right stage to be producing interstellar particles at Earth at the current epoch. 
Young main-sequence stars are potential sources of mm-sized particles at Earth, as they release material across a broad size range through the gravitational scattering of accumulating planetary solids. \cite{Murray2004}'s model yields a flux from $\beta$~Pic at Earth of $2 \times 10^{-9}$~km$^{-1}$~yr$^{-1}$ at diameters of 5 mm, and their total flux from similar systems (based on their Table 4) is three times this value. This total corresponds to three particles striking the Earth every year if their arrival directions are isotropic.

An alternative flux estimate is possible based on a simple extrapolation of the rate of appearance of km-scale interstellars. \cite{Hajdukova_2019} provide an estimate of 0.1 10~cm particles striking the Earth per year. This translates to $\sim 200$ events per year at 5~mm assuming the same canonical \citet{doh69} slope that they use.

Based on these estimates, we will adopt a fiducial expected rate of 3 to 200 events per year at 5~mm across the Earth's surface, or a flux of $6 \times 10^{-9}$ to $4 \times 10^{-7}$~km$^{-2}$ yr$^{-1}$. Given a collecting area of 17.5~million km~$^2$ and the efficiency factors given earlier, the probability that the portion of the GMN catalogue examined here would not contain any events given the expected rates is 97\% and 18\% respectively. Thus we can already conclude that the higher expected rate from \cite{hajstewie19} at 5~mm diameter is probably too high, either because the size distribution has a shallower slope than their assumed Dohnanyi value or due to the small number statistics inherent in extrapolating from a single event (1I/'Oumuamua).
Looking to the future, at the lower theoretical rate (the \cite{Murray2004} value for young main-sequence stars extrapolated to 5~mm diameter) and considering all events in the GMN catalog, then at its current collecting area GMN would have to observe for [25, 125] more years to be [50\%, 95\%] confident of observing at least one event. 
If GMN grows to cover all of the Earth's land mass (30\% of the surface) then it would have to observe for [3, 15] years to be [50\%, 95\%] confident of seeing at least one event at the \cite{Murray2004} rate. From these values we can see that the Global Meteor Network will certainly prove a very effective tool in constraining the true rates of interstellars at video sizes over the coming decade.

\section{Conclusions}
From a catalog of 2.18 million meteors from the Global Meteor Network spanning from late 2018 to late 2024, and after careful filtering, manual review, and careful data reduction to extract the highest quality events, we found no meteors with heliocentric orbits that were not consistent with being bound to the Sun. 

Interstellar meteors have high speeds relative to the Sun ($\gas$ 42 km/s) but strike the moving Earth from all sides, and so have a broad range of in-atmosphere speeds ($\sim$12 to 72 km/s). Our study focused only on meteors with geocentric velocities lower than 50~km~s$^{-1}$ to avoid the larger uncertainties associated with events with few video frames, and so to extract the highest quality events. This filter on meteor speed relative to the Earth does reduce our sensitivity to meteors with high speeds relative to the Sun arriving from the direction of the Earth's apex and these events will be examined more closely in future work.

We conclude that the flux of interstellar meteors at GMN visual sizes is consistent with a value less than $8 \pm 2 \times 10^{-11}$~km$^{-2}$~h$^{-1}$, and that fewer than 1 in $10^6$ optical meteors are interstellar. This flux limit is comparable to theoretical estimates of interstellar rates at these sizes, and the Global Meteor Network is well-positioned to place meaningful constraints on the flux of interstellars at video sizes over the next several years.

\section{acknowledgments}
This work was supported in part by NASA Meteoroid Environment Office under cooperative agreement 80NSSC24M0060 and by the Natural Sciences and Engineering Research Council of Canada (Grants no. RGPIN-2023-03538 \& RGPIN-2024-05200), and by the Canada Research Chairs Program. 

The authors would like to thank the following GMN contributors for making the data available for this work:
A. Campbell, Adam Mullins, Aden Walker, Adrian Bigland, Adriana Roggemans, Adriano Fonseca, Aksel Askanius, Alain Marin, Alaistar Brickhill, Alan Beech, Alan Maunder, Alan Pevec, Alan Pickwick, Alan Decamps, Alan Cowie, Alan Kirby, Alan Senior, Alastair Emerson, Aled Powell, Alejandro Barriuso, Aleksandar Merlak, Alex Bell, Alex Haislip, Alex Hodge, Alex Jeffery, Alex Kichev, Alex McConahay, Alex Pratt, Alex Roig, Alex Aitov, Alex McGuinness, Alexander Wiedekind-Klein, Alexander Kasten, Alexandre Alves, Alfredo Dal' Ava Júnior, Alison Scott, Amy Barron, Anatoly Ijon, Andre Rousseau, Andre Bruton, Andrea Storani, Andrei Marukhno, Andres Fernandez, Andrew Campbell-Laing, Andrew Challis, Andrew Cooper, Andrew Fiamingo, Andrew Heath, Andrew Moyle, Andrew Washington, Andrew Fulher, Andrew Robertson, Andy Stott, Andy Sapir, Andy Shanks, Ange Fox, Angel Sierra, Angélica López Olmos, Anna Johnston, Anoop Chemencherry, Ansgar Schmidt, Anthony Hopkinson, Anthony Pitt, Anthony Kesterton, Anton Macan, Anton Yanishevskiy, Antony Crowther, Anzhari Purnomo, Arie Blumenzweig, Arie Verveer, Arnaud Leroy, Arne Krueger, Attila Nemes, Barry Findley, Bart Dessoy, Bela Szomi Kralj, Ben Poulton, Bence Kiss, Bernard Côté, Bernard Hagen, Bev M. Ewen-Smith, Bill Cooke, Bill Wallace, Bill Witte, Bill Carr, Bill Thomas, Bill Kraimer, Bob Evans, Bob Greschke, Bob Hufnagel, Bob Marshall, Bob Massey, Bob Zarnke, Bob Guzik, Brenda Goodwill, Brendan Cooney, Brendon Reid, Brian Chapman, Brian Murphy, Brian Rowe, Brian Hochgurtel, Wyatt Hochgurtel, Brian Mitchell, Bruno Bonicontro, Bruno Casari, Callum Potter, Carl Elkins, Carl Mustoe, Carl Panter, Cesar Domingo Pardo, Charles Thody, Charlie McCormack, Chris Baddiley, Chris Blake, Chris Dakin, Chris George, Chris James, Chris Ramsay, Chris Reichelt, Chris Chad, Chris O'Neill, Chris White, Chris Jones, Chris Sale, Christian Wanlin, Christine Ord, Christof Zink, Christophe Demeautis, Christopher Coomber, Christopher Curtis, Christopher Tofts, Christopher Brooks, Chuck Goldsmith, Chuck Pullen, Ciaran Tangney, Claude Boivin, Claude Surprenant, Clive Sanders, Clive Hardy, Colin Graham, Colin Marshall, Colin Nichols, Con Stoitsis, Craig Young, Creina Beaman, Daknam Al-Ahmadi, Damien Lemay, Damien McNamara, Damir Matković, Damir Šegon, Damjan Nemarnik, Dan Klinglesmith, Dan Pye, Daniel Duarte, Daniel J. Grinkevich, Daniela Cardozo Mourão, Danijel Reponj, Danko Kočiš, Dario Zubović, Dave Jones, Dave Mowbray, Dave Newbury, Dave Smith, David Akerman, David Attreed, David Bailey, David Brash, David Castledine, David Hatton, David Leurquin, David Price, David Rankin, David Robinson, David Rollinson, David Strawford, David Taylor, David Rogers, David Banes, David Johnston, David Rees, David Cowan, David Greig, David Hickey, David Colthorpe, David Straer, David Harding, David Furneaux, Dean Moore, Debbie Godsiff, Denis Bergeron, Denis St-Gelais, Dennis Behan, Derek Poulton, Didier Walliang, Dimitris Georgoulas, Dino Čaljkušić, Dmitrii Rychkov, Dominique Guiot, Don Anderson, Don Hladiuk, Dorian Božičević, Dougal Matthews, Douglas Sloane, Douglas Stone, Dustin Rego, Dylan O’Donnell, Ed Breuer, Ed Harman, Edd Stone, Edgar Mendes Merizio, Edison José Felipe Pérezgómez Álvarez, Edson Valencia Morales, Eduardo Fernandez Del Peloso, Eduardo Lourenço, Edward Cooper, Egor Gustov, Ehud Behar, Eleanor Mayers, Emily Barraclough, Enrico Pettarin, Enrique Arce, Enrique Chávez Garcilazo, Eric Lopez, Eric Toops, Errol Balks, Erwin van Ballegoij, Erwin Harkink, Eugene Potapov, Ewan Richardson, Fabricio Borges, Fabricio Colvero, Fabrizio Guida, Ferenc-Levente Juhos, Fernando Dall'Igna, Fernando Jordan, Fernando Requena, Filip Matković, Filip Mezak, Filip Parag, Fiona Cole, Firuza Rahmat, Florent Benoit, Francis Rowsell, François Simard, Frank Lyter, Frantisek Bilek, Gabor Sule, Gaétan Laflamme, Gareth Brown, Gareth Lloyd, Gareth Oakey, Garry Dymond, Gary Parker, Gary Eason, Gavin Martin, Gene Mroz, Geoff Scott, Georges Attard, Georgi Momchilov, Gerard Van Os, Germano Soru, Gilberto Sousa, Gilton Cavallini, Gordon Hudson, Graeme Hanigan, Graeme McKay, Graham Stevens, Graham Winstanley, Graham Henstridge, Graham Atkinson, Graham Palmer, Graham Cann, Greg Michael, Greg Parker, Gulchehra Kokhirova, Gustav Frisholm, Gustavo Silveira B. Carvalho, Guy Létourneau, Guy Williamson, Guy Lesser, Hamish Barker, Hamish McKinnon, Haris Jeffrey, Harri Kiiskinen, Hartmut Leiting, Heather Petelo, Henning Letmade, Heriton Rocha, Hervé Lamy, Herve Roche, Holger Pedersen, Horst Meyerdierks, Howard Edin, Hugo González, Iain Drea, Ian Enting Graham, Ian Lauwerys, Ian Parker, Ian Pass, Ian A. Smith, Ian Williams, Ian Hepworth, Ian Collins, Igor Duchaj, Igor Henrique, Igor Macuka, Igor Pavletić, Ilya Jankowsky, Ioannis Kedros, Ivan Gašparić, Ivan Sardelić, Ivica Ćiković, Ivica Skokić, Ivo Dijan, Ivo Silvestri, Jack Barrett, Jacques Masson, Jacques Walliang, Jacqui Thompson, James Davenport, James Farrar, James Scott, James Stanley, James Dawson, Jamie Allen, Jamie Cooper, Jamie McCulloch, Jamie Olver, Jamie Shepherd, Jan Hykel, Jan Wisniewski, Janis Russell, Janusz Powazki, Jasminko Mulaomerović, Jason Burns, Jason Charles, Jason Gill, Jason van Hattum, Jason Sanders, Javor Kac, Jay Shaffer, Jean Francois Larouche, Jean Vallieres, Jean Brunet, Jean-Baptiste Kikwaya, Jean-Fabien Barrois, Jean-Louis Naudin, Jean-Marie Jacquart, Jean-Paul Dumoulin, Jean-Philippe Barrilliot, Jeff Holmes, Jeff Huddle, Jeff Wood, Jeff Devries, Jeffrey Legg, Jenifer Millard, Jeremy Taylor, Jeremy Spink, Jesse Stayte, Jesse Lard, Jessica Richards, Jim Blackhurst, Jim Cheetham, Jim Critchley, Jim Fordice, Jim Gilbert, Jim Rowe, Jim Seargeant, Jim Fakatselis, João Mattei, Joaquín Albarrán, Jochen Vollsted, Jocimar Justino, Joe Zender, John W. Briggs, John Drummond, John Hale, John Kmetz, John Maclean, John Savage, John Thurmond, John Tuckett, John Waller, John Wildridge, John Bailey, John Thompson, John Martin, John Yules, Jon Bursey, Jonathan Alexis Valdez Aguilar, Jonathan Eames, Jonathan Mackey, Jonathan Whiting, Jonathan Wyatt, Jonathon Kambulow, Jorge Augusto Acosta Bermúdez, Jorge Oliveira, Jose Carballada, Jose Galindo Lopez, José María García, José-Luis Martín, Josef Scarantino, Josip Belas, Josip Krpan, Jost Jahn, Juan Luis Muñoz, Juergen Neubert, Julián Martínez, Jure Zakrajšek, Jürgen Dörr, Jürgen Ketterer, Justin Zani, Karen Smith, Karl Browne, Kath Johnston, Kees Habraken, Keith Maslin, Keith Biggin, Keith Christie, Kelvin Richards, Ken Jamrogowicz, Ken Lawson, Ken Gledhill, Ken Kirvan, Ken Whitnall, Kevin Gibbs-Wragge, Kevin Morgan, Kevin Faure, Klaas Jobse, Korado Korlević, Kyle Francis, Lachlan Gilbert, Larry Groom, Laurent Brunetto, Laurie Stanton, Lawrence Saville, Lee Hill, Lee Brady, Leith Robertson, Len North, Les Rowe, Leslie Kaye, Lev Pustil’Nik, Lexie Wallace, Lisa Holstein, Llewellyn Cupido, Logan Carpenter, Lorna McCalman, Louw Ferreira, Lovro Pavletić, Lubomir Moravek, Luc Turbide, Luc Busquin, Lucia Dowling, Luciano Miguel Diniz, Ludger Börgerding, Luis Fabiano Fetter, Luis Santo, Maciej Reszelsk, Magda Wisniewska, Manel Colldecarrera, Marc Corretgé Gilart, Marcel Berger, Marcelo Domingues, Marcelo Zurita, Marcio Malacarne, Marco Verstraaten, Marcus Rigo, Margareta Gumilar, Marián Harnádek, Mariusz Adamczyk, Mark Fairfax, Mark Gatehouse, Mark Haworth, Mark McIntyre, Mark Phillips, Mark Robbins, Mark Spink, Mark Suhovecky, Mark Williams, Mark Ward, Mark Bingham, Mark Fechter, Marko Šegon, Marko Stipanov, Marshall Palmer, Marthinus Roos, Martin Breukers, Martin Richmond-Hardy, Martin Robinson, Martin Walker, Martin Woodward, Martin Connors, Martin Kobliha, Martyn Andrews, Mary Waddingham, Mary Hope, Mason McCormack, Mat Allan, Matej Mihelčić, Matt Cheselka, Matthew Howarth, Matthew Finch, Max Schmid, Megan Gialluca, Mia Boothroyd, Michael Cook, Michael Mazur, Michael O’Connell, Michael Krocil, Michael Camilleri, Michael Kennedy, Michael Lowe, Michael Atkinson, Michał Warchoł, Michel Saint-Laurent, Miguel Diaz Angel, Miguel Preciado, Mike Breimann, Mike Hutchings, Mike Read, Mike Shaw, Mike Ball, Mike Youmans, Milan Kalina, Milen Nankov, Miles Eddowes, Minesh Patel, Miranda Clare, Mirjana Malarić, Muhammad Luqmanul Hakim Muharam, Murray Forbes, Murray Singleton, Murray Thompson, Myron Valenta, Nalayini Brito, Nawaz Mahomed, Ned Smith, Nedeljko Mandić, Neil Graham, Neil Papworth, Neil Waters, Neil Petersen, Neil Allison, Nelson Moreira, Neville Vann, Nial Bruce, Nicholas Hill, Nicholas Ruffier, Nick Howarth, Nick James, Nick Moskovitz, Nick Norman, Nick Primavesi, Nick Quinn, Nick Russel, Nick Powell, Nick Wiffen, Nicola Masseroni, Nigel Bubb, Nigel Evans, Nigel Owen, Nigel Harris, Nigen Harris, Nikola Gotovac, Nikolay Gusev, Nikos Sioulas, Noah Simmonds, Norman Izsett, Olaf Jakubzik Reinartz, Ollie Eisman, Pablo Canedo, Pablo Domingo Escat, Paraksh Vankawala, Pat Devine, Patrick Franks, Patrick Poitevin, Patrick Geoffroy, Patrick Onesty, Patrik Kukić, Paul Cox, Paul Dickinson, Paul Haworth, Paul Heelis, Paul Kavanagh, Paul Ludick, Paul Prouse, Paul Pugh, Paul Roche, Paul Roggemans, Paul Stewart, Paul Huges, Paul Breck, Pedro Augusto de Jesus Castro, Penko Yordanov, Pete Graham, Pete Lynch, Peter G. Brown, Peter Campbell-Burns, Peter Davis, Peter Eschman, Peter Gural, Peter Hallett, Peter Jaquiery, Peter Kent, Peter Lee, Peter McKellar, Peter Meadows, Peter Stewart, Peter Triffitt, Peter Leigh, Peter Felhofer, Péter Molnár, Pető Zsolt, Phil James, Philip Gladstone, Philip Norton, Philippe Schaak, Phillip Wilhelm Maximilian Grammerstorf, Pierre Gamache, Pierre de Ponthière, Pierre-Michael Micaletti, Pierre-Yves Pechart, Pieter Dijkema, Predrag Vukovic, Przemek Nagański, Radim Stano, Rajko Sušanj, Raju Aryal, Ralph Brady, Raoul van Eijndhoven, Raul Truta, Raul Elias-Drago, Raymond Shaw, Rebecca Starkey, Reinhard Kühn, Remi Lacasse, Renato Cássio Poltronieri, René Tardif, Richard Abraham, Richard Bassom, Richard Croy, Richard Davis, Richard Fleet, Richard Hayler, Richard Johnston, Richard Kacerek, Richard Payne, Richard Stevenson, Richard Severn, Rick Fischer, Rick Hewett, Rick James, Ricky Bassom, Rob Agar, Rob de Corday Long, Rob Saunders, Rob Smeenk, Robert Longbottom, Robert McCoy, Robert Saint-Jean, Robert D. Steele, Robert Veronneau, Robert Peledie, Robert Haas, Robert Kiendl, Robert Vallone, Robin Boivin, Robin Earl, Robin Rowe, Robin Leadbeater, Roel Gloudemans, Roger Banks, Roger Morin, Roger Conway, Roland Idaczyk, Rolf Carstens, Roman Moryachkov, Romke Schievink, Romulo Jose, Ron James Jr, Ronal Kunkel, Roslina Hussain, Ross Skilton, Ross Dickie, Ross Welch, Ross Hortin, Russell Jackson, Russell Brunton, Ryan Frazer, Ryan Harper, Ryan Kinnett, Salvador Aguirre, Sam Green, Sam Hemmelgarn, Sam Leaske, Sarah Tonorio, Scott Kaufmann, Sebastian Klier, Seppe Canonaco, Seraphin Feller, Serge Bergeron, Sergio Mazzi, Sevo Nikolov, Simon Cooke-Willis, Simon Holbeche, Simon Maidment, Simon McMillan, Simon Minnican, Simon Parsons, Simon Saunders, Simon Fidler, Simon Oosterman, Simon Peterson, Simon lewis, Simon Lewis, Sofia Ulrich, Srivishal Sudharsan, Stacey Downton, Stan Nelson, Stanislav Korotkiy, Stanislav Tkachenko, Stef Vancampenhout, Stefan Frei, Stephane Zanoni, Stephen Grimes, Stephen Nattrass, Stephen M. Pereira, Steve Berry, Steve Bosley, Steve Carter, Steve Dearden, Steve Homer, Steve Kaufman, Steve Lamb, Steve Rau, Steve Tonkin, Steve Trone, Steve Welch, Steve Wyn-Harris, Steve Matheson, Steven Shanks, Steven Tilley, Stewart Doyle, Stewart Ball, Stuart Brett, Stuart Land, Stuart McAndrew, Sue Baker Wilson, Sylvain Cadieux, Tammo Jan Dijkema, Ted Cline, Terry Pundiak, Terry Richardson, Terry Simmich, Terry Young, Theodor Feldbaumer, Thiago Paes, Thilo Mies, Thomas Blog, Thomas Schmiereck, Thomas Stevenson, Thomas Duff, Tihomir Jakopčić, Tim Burgess, Tim Claydon, Tim Cooper, Tim Gloudemans, Tim Havens, Tim Polfliet, Tim Frye, Tioga Gulon, Tobias Westphal, Tom Warner, Tom Bell, Tommy McEwan, Tommy B. Nielsen, Torcuill Torrance, Tosh White, Tracey Snelus, Travis Shao, Trevor Clifton, Ubiratan Borges, Urs Wirthmueller, Uwe Glässner, Vasilii Savtchenko, Ventsislav Bodakov, Victor Acciari, Viktor Toth, Vincent McDermott, Vitor Jose Pereira, Vladimir Jovanović, Vladimir Vusanin, Waily Harim, Warley Souza, Warwick Latham, Washington Oliveira, Wayne Metcalf, Wenceslao Trujillo, William Perkin, William Schauff, William Stewart, William Harvey, William Hernandez, Wullie Mitchell, Yakov Tchenak, Yanislav Ivanov, Yfore Scott, Yohsuke Akamatsu, Yong-Ik Byun, Yozhi Nasvadi, Yuri Stepanychev, Zach Steele, Zané Smit, Zbigniew Krzeminski, Željko Andreić, Zhuoyang Chen, Zoran Dragić, Zoran Knez, Zoran Novak, Zouhair Benkhaldoun, Asociación de Astronomía de Marina Alta, Costa Blanca Astronomical Society, Perth Observatory Volunteer Group, Phillips Academy Andover, Royal Astronomical Society of Canada Calgary Centre

%\end{acknowledgments}

%% To help institutions obtain information on the effectiveness of their 
%% telescopes the AAS Journals has created a group of keywords for telescope 
%% facilities.
%
%% Following the acknowledgments section, use the following syntax and the
%% \facility{} or \facilities{} macros to list the keywords of facilities used 
%% in the research for the paper.  Each keyword is check against the master 
%% list during copy editing.  Individual instruments can be provided in 
%% parentheses, after the keyword, but they are not verified.

\vspace{5mm}
\facilities{Global Meteor Network}

%% Similar to \facility{}, there is the optional \software command to allow 
%% authors a place to specify which programs were used during the creation of 
%% the manuscript. Authors should list each code and include either a
%% citation or url to the code inside ()s when available.

\software{WesternMeteorPyLib, Shapely, pykml, cartopy, pyproj}

%\appendix

%% Appendix material should be preceded with a single \appendix command.
%% There should be a \section command for each appendix. Mark appendix
%% subsections with the same markup you use in the main body of the paper.

%% Each Appendix (indicated with \section) will be lettered A, B, C, etc.
%% The equation counter will reset when it encounters the \appendix
%% command and will number appendix equations (A1), (A2), etc. The
%% Figure and Table counter will not reset.

\bibliography{GMN-interstellar}{}

\begin{thebibliography}{}
\expandafter\ifx\csname natexlab\endcsname\relax\def\natexlab#1{#1}\fi
\providecommand{\url}[1]{\href{#1}{#1}}
\providecommand{\dodoi}[1]{doi:~\href{http://doi.org/#1}{\nolinkurl{#1}}}
\providecommand{\doeprint}[1]{\href{http://ascl.net/#1}{\nolinkurl{http://ascl.net/#1}}}
\providecommand{\doarXiv}[1]{\href{https://arxiv.org/abs/#1}{\nolinkurl{https://arxiv.org/abs/#1}}}

\bibitem[{{Almond} {et~al.}(1950){Almond}, {Davies}, \& {Lovell}}]{Almond_1950}
{Almond}, M., {Davies}, J.~G., \& {Lovell}, A.~C.~B. 1950, The Observatory, 70,
  112

\bibitem[{Baggaley(2000)}]{Baggaley_2000}
Baggaley, W. 2000, … of Geophysical Research: Space Physics (1978 …, 105.
\newblock \url{http://onlinelibrary.wiley.com/doi/10.1029/1999JA900383/full}

\bibitem[{{Baggaley}(2000)}]{bag00}
{Baggaley}, W.~J. 2000, J. Geophys. Res., 105, 10353,
  \dodoi{10.1029/1999JA900383}

\bibitem[{{Baggaley} {et~al.}(1993){Baggaley}, {Taylor}, \&
  {Steel}}]{bagtayste93}
{Baggaley}, W.~J., {Taylor}, A.~D., \& {Steel}, D.~I. 1993, in Meteoroids and
  their Parent Bodies, ed. J.~{Stohl} \& I.~P. {Williams}, 53

\bibitem[{{Baguhl} {et~al.}(1996){Baguhl}, {Gr{\"u}n}, \&
  {Landgraf}}]{baggrulan96}
{Baguhl}, M., {Gr{\"u}n}, E., \& {Landgraf}, M. 1996, \ssr, 78, 165,
  \dodoi{10.1007/BF00170803}

\bibitem[{Baguhl {et~al.}(1995)Baguhl, Grün, Hamilton, Linkert, Riemann,
  Staubach, \& Zook}]{Baguhl1995}
Baguhl, M., Grün, E., Hamilton, D.~P., {et~al.} 1995, in The High Latitude
  Heliosphere, ed. R.~G. Marsden (Dordrecht: Springer Netherlands), 471–476,
  \dodoi{10.1007/978-94-011-0167-7_77}

\bibitem[{Binney \& Tremaine(1987)}]{bintre87}
Binney, J., \& Tremaine, S. 1987, Galactic Dynamics (Princeton: Princeton
  University Press)

\bibitem[{{Blaauw} {et~al.}(2016){Blaauw}, {Campbell-Brown}, \&
  {Kingery}}]{blacamkin16}
{Blaauw}, R.~C., {Campbell-Brown}, M., \& {Kingery}, A. 2016, \mnras, 463, 441,
  \dodoi{10.1093/mnras/stw1979}

\bibitem[{{Cooke} {et~al.}(1993){Cooke}, {Mulholland}, \&
  {Oliver}}]{coomuloli93}
{Cooke}, W.~J., {Mulholland}, J.~D., \& {Oliver}, J.~P. 1993, Advances in Space
  Research, 13, 119, \dodoi{10.1016/0273-1177(93)90577-X}

\bibitem[{{Delhaye}(1965)}]{del65}
{Delhaye}, J. 1965, in Galactic structure, ed. A.~{Blaauw} \& M.~{Schmidt}
  (University of Chicago Press), 61

\bibitem[{{Dohnanyi}(1969)}]{doh69}
{Dohnanyi}, J.~S. 1969, J. Geophys. Res., 74, 2531.
\newblock
  \url{https://agupubs.onlinelibrary.wiley.com/doi/pdfdirect/10.1029/JB074i010p02531}

\bibitem[{{Fisher}(1928)}]{fis28}
{Fisher}, W.~J. 1928, Harvard College Observatory Circular, 331, 1

\bibitem[{Froncisz {et~al.}(2020)Froncisz, Brown, \& Weryk}]{Froncisz_2020}
Froncisz, M., Brown, P., \& Weryk, R.~J. 2020, Planetary and Space Science,
  190, 104980, \dodoi{10.1016/j.pss.2020.104980}

\bibitem[{{Gehrels}(1986)}]{geh86}
{Gehrels}, N. 1986, \apj, 303, 336, \dodoi{10.1086/164079}

\bibitem[{Gillies {et~al.}(2024)Gillies, van~der Wel, Van~den Bossche, Taves,
  Arnott, Ward, {et~al.}}]{gillies_2024_13345370}
Gillies, S., van~der Wel, C., Van~den Bossche, J., {et~al.} 2024, Shapely,
  2.0.6,  Zenodo, \dodoi{10.5281/zenodo.13345370}

\bibitem[{{{Gr{\"u}n}} {et~al.}(1994){{Gr{\"u}n}}, {Gustafson}, {Mann},
  {Baguhl}, {Morfill}, {Staubach}, {Taylor}, \& {Zook}}]{grugusman94}
{{Gr{\"u}n}}, E., {Gustafson}, B., {Mann}, I., {et~al.} 1994, A\&A, 286, 915

\bibitem[{{Gr{\"u}n} {et~al.}(2000){Gr{\"u}n}, {Landgraf}, {Hor{\'a}nyi},
  {Kissel}, {Kr{\"u}ger}, {Srama}, {Svedhem}, \& {Withnell}}]{grulanhor00}
{Gr{\"u}n}, E., {Landgraf}, M., {Hor{\'a}nyi}, M., {et~al.} 2000, \jgr, 105,
  10403, \dodoi{10.1029/1999JA900376}

\bibitem[{{Gr{\"u}n} {et~al.}(1997){Gr{\"u}n}, {Staubach}, {Baguhl},
  {Hamilton}, {Zook}, {Dermott}, {Gustafson}, {Fechtig}, {Kissel}, {Linkert},
  {Linkert}, {Srama}, {Hanner}, {Polanskey}, {Horanyi}, {Lindblad}, {Mann},
  {McDonnell}, {Morfill}, \& {Schwehm}}]{grustabag97}
{Gr{\"u}n}, E., {Staubach}, P., {Baguhl}, M., {et~al.} 1997, \icarus, 129, 270,
  \dodoi{10.1006/icar.1997.5789}

\bibitem[{{Hajdukov{\'a}} {et~al.}(2019){Hajdukov{\'a}}, {Sterken}, \&
  {Wiegert}}]{hajstewie19}
{Hajdukov{\'a}}, M{\'a}ria, J., {Sterken}, V., \& {Wiegert}, P. 2019, in
  Meteoroids: Sources of Meteors on Earth and Beyond, ed. G.~O. {Ryabova},
  D.~J. {Asher}, \& M.~J. {Campbell-Brown} (Cambridge University Press), 235

\bibitem[{{Hajdukova}(1994)}]{haj94}
{Hajdukova}, M. 1994, Astronomy and Astrophysics, 288, 330

\bibitem[{Hajdukova {et~al.}(2014)Hajdukova, Kornoš, \&
  Tóth}]{Hajdukova__2014}
Hajdukova, M., Kornoš, L., \& Tóth, J. 2014, Meteoritics \& Planetary
  Science, 49, 63–68, \dodoi{10.1111/maps.12119}

\bibitem[{{Hajdukova} {et~al.}(2020){Hajdukova}, {Sterken}, {Wiegert}, \&
  {Korno{\v{s}}}}]{hajstewie20}
{Hajdukova}, M., {Sterken}, V., {Wiegert}, P., \& {Korno{\v{s}}}, L. 2020,
  \planss, 192, 105060, \dodoi{10.1016/j.pss.2020.105060}

\bibitem[{{Hajdukov{\'a}} {et~al.}(2024){Hajdukov{\'a}}, {Stober}, {Barghini},
  {Koten}, {Vaubaillon}, {Sterken}, {{\v{D}}uri{\v{s}}ov{\'a}}, {Jackson}, \&
  {Desch}}]{hajstobar24}
{Hajdukov{\'a}}, M., {Stober}, G., {Barghini}, D., {et~al.} 2024, \aap, 691,
  A8, \dodoi{10.1051/0004-6361/202449569}

\bibitem[{{Hajdukov{\'a}} \& {Paulech}(2002)}]{hajpau02}
{Hajdukov{\'a}}, Jr., M., \& {Paulech}, T. 2002, in ESA Special Publication,
  Vol. 500, Asteroids, Comets, and Meteors: ACM 2002, ed. {B.~Warmbein},
  173--176

\bibitem[{{Hajdukov{\'a}} \& {Paulech}(2007)}]{hajpau07}
{Hajdukov{\'a}}, Jr., M., \& {Paulech}, T. 2007, Contributions of the
  Astronomical Observatory Skalnate Pleso, 37, 18

\bibitem[{Hajduková(1994)}]{Hajdukova_1994}
Hajduková. 1994, Astronomy and Astrophysics, 288, 330–334

\bibitem[{Hajduková {et~al.}(2019)Hajduková, Sterken, \&
  Wiegert}]{Hajdukova_2019}
Hajduková, M., Sterken, V., \& Wiegert, P. 2019, in Meteoroids: Sources of
  Meteors on Earth and Beyond, ed. R.~G. O., A.~D. J., \& C.-B.~M. D (Cambridge
  University Press), 235

\bibitem[{{Hawkes} {et~al.}(1999){Hawkes}, {Close}, \&
  {Woodworth}}]{hawclowoo99}
{Hawkes}, R.~L., {Close}, T., \& {Woodworth}, S. 1999, in Meteroids 1998, ed.
  W.~J. {Baggaley} \& V.~{Porubcan}, 257

\bibitem[{{H{\"o}fner}(2008)}]{hof08}
{H{\"o}fner}, S. 2008, \aap, 491, L1, \dodoi{10.1051/0004-6361:200810641}

\bibitem[{Jaschek \& Valbousquet(1992)}]{jaschek_solar_1992}
Jaschek, C., \& Valbousquet, A. 1992, Astronomy and Astrophysics, 255, 124.
\newblock \url{https://ui.adsabs.harvard.edu/abs/1992A&A...255..124J}

\bibitem[{King {et~al.}(2013)King, Platnick, Menzel, Ackerman, \&
  Hubanks}]{king2013spatial}
King, M.~D., Platnick, S., Menzel, W.~P., Ackerman, S.~A., \& Hubanks, P.~A.
  2013, IEEE transactions on geoscience and remote sensing, 51, 3826

\bibitem[{{Kresak} \& {Kresakova}(1976)}]{krekre76}
{Kresak}, L., \& {Kresakova}, M. 1976, Bulletin of the Astronomical Institutes
  of Czechoslovakia, 27, 106

\bibitem[{{Landgraf} {et~al.}(2000){Landgraf}, {Baggaley}, {Gr{\"u}n},
  {Kr{\"u}ger}, \& {Linkert}}]{lanbaggru00}
{Landgraf}, M., {Baggaley}, W.~J., {Gr{\"u}n}, E., {Kr{\"u}ger}, H., \&
  {Linkert}, G. 2000, J. Geophys. Res., 105, 10343,
  \dodoi{10.1029/1999JA900359}

\bibitem[{{Landgraf} \& {Gr{\"u}n}(1998)}]{langru98}
{Landgraf}, M., \& {Gr{\"u}n}, E. 1998, in IAU Colloq. 166: The Local Bubble
  and Beyond, ed. D.~{Breitschwerdt}, M.~J. {Freyberg}, \& J.~{Truemper}, Vol.
  506 (Springer), 381--384, \dodoi{10.1007/BFb0104750}

\bibitem[{Lovell(1954)}]{lov54}
Lovell, A. C.~B. 1954, Meteor Astronomy (Oxford: Clarendon).
\newblock \url{https://ui.adsabs.harvard.edu/abs/1954meas.book.....L}

\bibitem[{{Mathews} {et~al.}(1999){Mathews}, {Meisel}, {Janches}, {Getman}, \&
  {Zhou}}]{matmeijan99}
{Mathews}, J.~D., {Meisel}, D.~D., {Janches}, D., {Getman}, V.~S., \& {Zhou},
  Q.~H. 1999, in Meteroids 1998, ed. W.~J. {Baggaley} \& V.~{Porubcan}, 79

\bibitem[{{Mattsson} \& {H{\"o}fner}(2011)}]{mathof11}
{Mattsson}, L., \& {H{\"o}fner}, S. 2011, \aap, 533, A42,
  \dodoi{10.1051/0004-6361/201015572}

\bibitem[{McKinley(1961)}]{McKinley_1961}
McKinley, D. 1961, Meteor Science and Engineering (McGraw-Hill)

\bibitem[{Meech {et~al.}(2017)Meech, Weryk, Micheli, Kleyna, Hainaut, Jedicke,
  Wainscoat, Chambers, Keane, Petric, Denneau, Magnier, Berger, Huber,
  Flewelling, Waters, Schunová-Lilly, \& Chastel}]{Meech_2017}
Meech, K.~J., Weryk, R.~J., Micheli, M., {et~al.} 2017, Nature, 1–12,
  \dodoi{10.1038/nature25020}

\bibitem[{{Meeker} {et~al.}(2017){Meeker}, {Hahn}, \& {Escobar}}]{meehahesc17}
{Meeker}, W.~Q., {Hahn}, G.~J., \& {Escobar}, L.~A. 2017, Statistical Intervals
  for a Poisson Distribution (John Wiley \& Sons, Ltd), 149--161,
  \dodoi{https://doi.org/10.1002/9781118594841.ch7}

\bibitem[{{Meisel} {et~al.}(2002{\natexlab{a}}){Meisel}, {Janches}, \&
  {Mathews}}]{meijanmat02a}
{Meisel}, D.~D., {Janches}, D., \& {Mathews}, J.~D. 2002{\natexlab{a}}, ApJ,
  567, 323, \dodoi{10.1086/322317}

\bibitem[{{Meisel} {et~al.}(2002{\natexlab{b}}){Meisel}, {Janches}, \&
  {Mathews}}]{meijanmat02b}
---. 2002{\natexlab{b}}, ApJ, 579, 895, \dodoi{10.1086/342919}

\bibitem[{Mihalas \& Binney(1981)}]{mihbin81}
Mihalas, D., \& Binney, J. 1981, Galactic Astronomy (New York: W. H. Freeman
  and Co.)

\bibitem[{Murray {et~al.}(2004)Murray, Weingartner, \& Capobianco}]{Murray2004}
Murray, N., Weingartner, J.~C., \& Capobianco, C. 2004, Astrophysical Journal,
  600, 804–827

\bibitem[{Musci {et~al.}(2012)Musci, Weryk, Brown, Campbell-Brown, \&
  Wiegert}]{Musci_2012}
Musci, R., Weryk, R.~J., Brown, P.~G., Campbell-Brown, M., \& Wiegert, P. 2012,
  The Astrophysical Journal, 745, 161–167,
  \dodoi{10.1088/0004-637X/745/2/161}

\bibitem[{Opik(1950)}]{Opik_1950}
Opik, E.~J. 1950, Irish Astronomical Journal, 1, 80

\bibitem[{Reid \& Brunthaler(2004)}]{Reid_2004}
Reid, M.~J., \& Brunthaler, A. 2004, The Astrophysical Journal, 616, 872

\bibitem[{Snow {et~al.}(2024)Snow, Whitaker, Cochran, Miara, den Bossche, Mayo,
  Lucas, Cochrane, de~Kloe, Karney, Shaw, Filipe, Ouzounoudis, Dearing, Lostis,
  Hoese, Couwenberg, Jurd, Gohlke, Schneck, McDonald, Itkin, May,
  de~Bittencourt, Little, Rahul, Eubank, Neil, \&
  Taves}]{alan_d_snow_2024_13864781}
Snow, A.~D., Whitaker, J., Cochran, M., {et~al.} 2024, pyproj4/pyproj: 3.7.0
  Release, 3.7.0,  Zenodo, \dodoi{10.5281/zenodo.13864781}

\bibitem[{{SonotaCo}(2009)}]{son09}
{SonotaCo}. 2009, WGN, Journal of the International Meteor Organization, 37, 55

\bibitem[{Sterken {et~al.}(2012)Sterken, Altobelli, Kempf, Schwehm, Srama, \&
  Grün}]{sterken_flow_2012}
Sterken, V.~J., Altobelli, N., Kempf, S., {et~al.} 2012, Astronomy \&
  Astrophysics, 538, A102, \dodoi{10.1051/0004-6361/201117119}

\bibitem[{Stohl(1970)}]{Stohl_1970}
Stohl, J. 1970, Bulletin of the Astronomical Institutes of …, 21, 10–17

\bibitem[{{Taylor} {et~al.}(1996){Taylor}, {Baggaley}, \&
  {Steel}}]{taybagste96}
{Taylor}, A.~D., {Baggaley}, W.~J., \& {Steel}, D.~I. 1996, \nat, 380, 323,
  \dodoi{10.1038/380323a0}

\bibitem[{Van de Sande {et~al.}(2020)Van de Sande, Walsh, \&
  Danilovich}]{vanwaldan20}
Van de Sande, M., Walsh, C., \& Danilovich, T. 2020, Monthly Notices of the
  Royal Astronomical Society, 495, 1650, \dodoi{10.1093/mnras/staa1270}

\bibitem[{Vida {et~al.}(2020{\natexlab{a}})Vida, Brown, Campbell-Brown,
  Wiegert, \& Gural}]{vida2020results}
Vida, D., Brown, P.~G., Campbell-Brown, M., Wiegert, P., \& Gural, P.~S.
  2020{\natexlab{a}}, Monthly Notices of the Royal Astronomical Society, 491,
  3996, \dodoi{10.1093/mnras/stz3338}

\bibitem[{Vida {et~al.}(2020{\natexlab{b}})Vida, Campbell-Brown, Brown, Egal,
  \& Mazur}]{vida2020draconids}
Vida, D., Campbell-Brown, M., Brown, P.~G., Egal, A., \& Mazur, M.~J.
  2020{\natexlab{b}}, Astronomy \& Astrophysics, 635, A153,
  \dodoi{10.1051/0004-6361/201937296}

\bibitem[{Vida {et~al.}(2020{\natexlab{c}})Vida, Gural, Brown, Campbell-Brown,
  \& Wiegert}]{vida2020theory}
Vida, D., Gural, P.~S., Brown, P.~G., Campbell-Brown, M., \& Wiegert, P.
  2020{\natexlab{c}}, Monthly Notices of the Royal Astronomical Society, 491,
  2688, \dodoi{10.1093/mnras/stz3160}

\bibitem[{{Vida} {et~al.}(2016){Vida}, {Zubovi{\'c}}, {{\v{S}}egon}, {Gural},
  \& {Cupec}}]{vidzubseg16}
{Vida}, D., {Zubovi{\'c}}, D., {{\v{S}}egon}, D., {Gural}, P., \& {Cupec}, R.
  2016, in International Meteor Conference Egmond, the Netherlands, 2-5 June
  2016, ed. A.~{Roggemans} \& P.~{Roggemans}, 307

\bibitem[{{Vida} {et~al.}(2021){Vida}, {{\v{S}}egon}, {Gural}, {Brown},
  {McIntyre}, {Dijkema}, {Pavleti{\'c}}, {Kuki{\'c}}, {Mazur}, {Eschman},
  {Roggemans}, {Merlak}, \& {Zubovi{\'c}}}]{vidseggur21}
{Vida}, D., {{\v{S}}egon}, D., {Gural}, P.~S., {et~al.} 2021, \mnras, 506,
  5046, \dodoi{10.1093/mnras/stab2008}

\bibitem[{{Von Nie{\ss}l} \& {Hoffmeister}(1925)}]{vonhof25}
{Von Nie{\ss}l}, G., \& {Hoffmeister}, C. 1925, {Katalog der
  Bestimmungsgr\"o{\ss}en f\"ur 611 Bahnen gro{\ss}er Meteore}, Vol. 100 (Wien:
  Denkschriften der Akademie der Wissenschaft.
  Mathematisch-Naturwissenchaftliche Klasse)

\bibitem[{{Wehry} \& {Mann}(1999)}]{wehman99}
{Wehry}, A., \& {Mann}, I. 1999, \aap, 341, 296

\bibitem[{{Weryk} \& {Brown}(2004)}]{werbro04}
{Weryk}, R.~J., \& {Brown}, P. 2004, Earth Moon and Planets, 95, 221,
  \dodoi{10.1007/s11038-005-9034-x}

\bibitem[{{Wiegert}(2014)}]{wie14}
{Wiegert}, P.~A. 2014, Icarus, 242, 112, \dodoi{10.1016/j.icarus.2014.06.031}

\bibitem[{{Zook} \& {Berg}(1975)}]{zoober75}
{Zook}, H.~A., \& {Berg}, O.~E. 1975, Plan. Space Sci., 23, 183,
  \dodoi{10.1016/0032-0633(75)90078-1}

\end{thebibliography}
\bibliographystyle{aasjournal}

%% This command is needed to show the entire author+affiliation list when
%% the collaboration and author truncation commands are used.  It has to
%% go at the end of the manuscript.
%\allauthors

%% Include this line if you are using the \added, \replaced, \deleted
%% commands to see a summary list of all changes at the end of the article.
%\listofchanges

\end{document}